\documentclass{aa}  

\usepackage{graphicx}
\graphicspath{{Figures/}}
\usepackage{txfonts}
\usepackage{float}
\usepackage{hyperref}
\usepackage{color}
\usepackage[usenames,dvipsnames]{xcolor}
\newcommand{\cla}[1]{#1}

\begin{document}

   \title{Magnetic field of the eclipsing binary UV Piscium}

   \subtitle{}

   \author{A. Hahlin\inst{1}
          \and
          O. Kochukhov\inst{1}
          \and
          E. Alecian\inst{2}
          \and
          J. Morin\inst{3}
          \and
          the BinaMIcS collaboration
          }

   \institute{Department of Physics and Astronomy, Uppsala University, Box 516, SE-751 20 Uppsala, Sweden\\\email{axel.hahlin@physics.uu.se}
   \and          
   Universit\'e Grenoble Alpes, CNRS, IPAG, F-38000 Grenoble, France
   \and
   LUPM, Universit\'e de Montpellier, CNRS, Place Eug\`ene Bataillon, F-34095 Montpellier, France}

   \date{Received 19 Mars 2021; accepted 1 May 2021}

 
  \abstract
   {}
   {The goal of this work is to study magnetic fields of the cool, eclipsing binary star UV Piscium (UV Psc). This system contains two active late-type stars, UV Psc A (G5V) and B (K3V). To obtain a complete picture, the properties of both global and local magnetic field structures are studied for both components.}
   {High-resolution intensity and circular polarisation spectra, collected in 2016 with the ESPaDOnS spectropolarimeter at the CFHT, were used to analyse the magnetic field of UV Psc. To increase the signal-to-noise ratio, the multi-line technique of least-squares deconvolution (LSD) was used to obtain average Stokes $IV$ profiles. Then, a Zeeman-Doppler imaging (ZDI) code 
   was employed to obtain the large-scale magnetic field topology and brightness distribution for both components of UV Psc. In addition, the small-scale magnetic fields, not visible to ZDI, were studied using the Zeeman intensification of \ion{Fe}{i} lines.
   }
   {The orbital and fundamental parameters of the system were revised based on the new radial velocity measurements.
   Maps of the surface magnetic field for both components of UV Psc were obtained, the large-scale magnetic fields feature strong toroidal and non-axisymetric components. UV Psc A and B have average global field strengths of 137~G and 88~G, respectively. The small-scale fields are notably stronger, with average strengths of 2.5 and 2.2 kG, respectively. Only $\sim$\,5\% of the total magnetic field strength is recovered with ZDI. Our results are in agreement with previous studies of partly-convective stars. Overall, UV Psc A has a stronger magnetic field compared to UV Psc B. Due to the eclipsing binary geometry, certain magnetic field features are not detectable using circular polarisation only. An investigation into theoretical linear polarisation profiles shows that they could be used to reveal antisymmetric components of the magnetic field. This result also has implications for the study of exoplanetary transit hosts. The successful use of Zeeman intensification shows the method's ability to extract information on magnetic field\cla{s} for stars rotating significantly more rapidly than in previous studies.}
   {}


   \maketitle
%

\section{Introduction}
Magnetic fields are known to play an important role in stellar evolution. They are key for slowing down the rotation rate of stars \citep[see e.g.][]{matt:2015}. \cla{The stellar wind is also affected by the magnetic fields as the wind structure is shaped by extended magnetic field lines \citep[e.g.][]{fionnagain:2019}. In addition, the wind speed has been associated with the magnetic topology of the origin area of the stellar wind \citep[e.g.][]{Cranmer:2019}
and the wind acceleration mechanism is linked to Alfv\'en wave turbulence dissipation \citep{holst:2014}.} They are also responsible for the emission of high energy particles and radiation. Many of these phenomena will also affect the environment around the star, potentially including planetary systems \citep[see e.g.][]{gallet:2017}. This will have a significant impact on any planetary atmosphere. Although magnetic fields are known to affect stars and their surroundings, they are often omitted during modelling of processes where it could have an effect on the outcome. The study of magnetic fields on and around different types of stars could help to increase the understanding of such processes and further improve stellar models.

A popular method to obtain a measure of the magnetic flux of a star is to observe its X-ray luminosity. A power law connecting the X-ray luminosity with the magnetic flux was established by \cite{pevtsov:2003}. While useful when no magnetic measurements are available, its accuracy might not be sufficient for meaningful estimates of magnetic fields for individual stars \citep{kochukhov:2019}. Regardless of accuracy, observations of signals directly originating from magnetic fields on the stellar surface are preferred. For this, the Zeeman effect \citep{donati:2009} is a powerful tool. By studying the effect of magnetic fields on spectral lines, information on both the strength and topology of \cla{the} surface magnetic field can be obtained.

To obtain a detailed understanding of the magnetic fields on the surface of a star, multiple spatial scales need to be studied. Since few stars can be resolved with interferometry, the methods used must obtain spatial information through indirect means. A common technique to extract information about the magnetic field structure is Zeeman-Doppler imaging \citep[ZDI, see e.g.][]{kochukhov:2016}. This method combines information on how the magnetic field affects spectral lines with the Doppler shift caused by stellar rotation and rotational modulation. By using polarised spectra, a global magnetic field map can be reconstructed. Mapping the large-scale magnetic component of the field is particularly important for understanding \cla{the} interaction of the star with its environment. This method does not give a complete picture of the magnetic fields on stars however. Due to the cancellation of polarisation signals corresponding to opposite field polarities, small-scale features are not visible in the disk-integrated polarised spectra. This has been demonstrated, for example, by simulating polarisation signatures of \cla{the} Sun as a star from fully resolved vector magnetograms \citep{kochukhov:2017a} or using theoretical high-resolution field maps resulting from flux transport simulations \citep{lehmann:2019}. The loss of the small-scale magnetic fields is an important issue since over 90\% of the magnetic energy has been found to be contained within these small scales in most cool stars \citep{reiners:2012,kochukhov:2021}. To probe these tangled, small-scale magnetic fields, this study also makes use of Zeeman intensification \citep[e.g.][]{kochukhov:2020a}. This method is not sensitive to the direction of the magnetic field and for this reason it does not suffer from the signal cancellation of nearby surface elements of an opposite polarity affecting polarisation observables.

This study investigates both the global and small-scale magnetic fields of the short-period eclipsing binary UV Piscium (UV Psc, HD\,7700, HIP\,5980). Eclipsing binaries are particularly interesting because their masses and radii can be obtained from observations, without assumptions about stellar models. While single stars might have uncertainties of up to 10\% for these parameters, analysis of binary stars can routinely reach errors of $\sim$ 1\% in mass. Furthermore, only for eclipsing binaries can the radii be determined with a similar accuracy. To asses the validity of stellar models, these accuracies are necessary to ensure sufficiently strong constraints in order to discard incorrect models \citep{torres:2010}. Another advantage of studying binary systems is that they have formed together \cla{\citep{Raghavan:2010}}. This means that they have the same age and their evolution has started from the same initial conditions. The effects of the magnetic field can therefore be more reliably disentangled from other processes affecting the star. 

The effect\cla{s} of stellar activity on the parameters of UV Psc have been studied before, a direct investigation of its magnetic field is therefore of particular interest. When \cite{popper:1997} studied this binary system, he was unable to fit the two components to the same isochrone. UV Psc A's age was estimated to be 9 Gyr, while the age of UV Psc B was estimated to be more than 20 Gyr. These results are not compatible with the standard stellar formation scenario that expects close binary components to originate from \cla{a} common protostellar cloud at the same time. Their estimated ages should therefore be similar. Since this discovery, \cite{ribas:2006} and \cite{lopez-morales:2007} have connected underestimation of stellar radii with magnetic activity in active binaries. Compared to conventional stellar models, rapidly rotating, active binaries have observed radii that are inflated by about 5--15\% \citep{ribas:2006}. \cla{The} components of UV Psc are known to rotate very fast ($P_{\rm rot}=P_{\rm orb}=0.86$~d) and therefore are magnetically active, which could possibly explain the age discrepancy. This problem was investigated by \cite{feiden:2013}. By including magnetic fields in stellar evolution models, consistent component ages of 4.7 Gyr were derived. Magnetic field strengths of 2.0~kG and 4.6~kG for UV Psc A and B respectively were needed to achieve this. While other parameters of this theoretical model could affect the result and final magnetic field strengths, it highlights that magnetic fields should not be neglected in stellar modelling. Accurate empirical magnetic field measurements are needed to validate theoretical results and further improve the modelling of active stars.

A special challenge that presents itself when studying UV Psc is that its components are rapidly rotating stars. Whereas such objects are optimal for application of ZDI, Zeeman intensification studies have not been performed for such rapidly rotating stars so far. Blending and broadening causes issues in finding spectral lines that can be used to effectively obtain information on the surface magnetic field. A successful use of Zeeman intensification would imply that the method is capable of yielding results at high rotation rates. This would open up the possibility of studying the surface magnetic fields of other rapidly rotating stars.

The rest of this paper is organised as follows.
In Sect.~\ref{ch:obs}, the observations of UV Psc that are used to analyse its magnetic fields are presented. Sections \ref{sec:LSD} and \ref{sec:SpecOrb} cover the data analysis to produce spectra that are usable for ZDI and Zeeman intensification study. Section \ref{sec:SpecOrb} also describes the revised orbital solution of UV Psc. In Sect.~\ref{sec:ZDI}, the global magnetic field topology is derived with the ZDI technique. Section \ref{ch:ZIntensification} is dedicated to the analysis of small-scale magnetic field\cla{s}, detectable through Zeeman intensification. Finally, Sect.~\ref{sec:summary} concludes the paper by summarising the results and comparing them with previous findings.

\section{Observations}
\label{ch:obs}
High-resolution spectropolarimetric observations of UV Psc were obtained with the ESPaDOnS spectropolarimeter \cla{\citep{donati:2003,wade:2016}} at the Canada-France-Hawaii Telescope (CHFT) in the context of the \cla{Binarity and Magnetic Interactions in various classes of stars (BinaMIcS)} large programme \citep{alecian:2015}. ESPaDOnS allows intensity spectra to be recorded, at a resolution of $R=65000$, simultaneously with either circular or linear polarisation. The observations analysed here were carried out in two intervals, the first, comprising two observations, between August 2 and 5, 2016, followed by another 16 observations between September 17 and 21 of the same year. The observations covered the wavelength range 3700--10000~\AA. In total, 18 sequences of 4$\times$450~s exposures were obtained in circular polarisation. Sub-exposures in each sequence were obtained with different polarimeter configurations in order to mitigate spurious polarisation \citep{donati:1997}. For the intensity spectra, each sub-exposure can be used as an independent observation, yielding a total of 72 Stokes $I$ spectra. No observations of linear polarisation have been made for UV Psc, \cla{so} the study of the global magnetic field topology is therefore limited to the intensity and circular polarisation spectra only.

The spectra used for the analysis of UV Psc have been reduced by the UPENA pipeline running the LIBRE-ESpRIT software \citep{donati:1997}. The continuum normalisation was carried out with the code described by \citet{rosen:2018}. The median signal-to-noise ratio (S/N) of the reduced Stokes $I$ spectra of UV Psc was determined to be 155 around the wavelength 5500~\AA. For Stokes $V$ the median S/N is 315 around the same wavelength. A log of all obtained Stokes $V$ measurements can be seen in Table~\ref{tab:StokesV}, including the mid-exposure Julian dates and orbital phases.

An important issue that has to be considered when interpreting polarisation observations of UV Psc is that the observation time might not be negligible compared to the orbital period of the system. With an orbital period of 0.86~d, the total time covered by the four sub-exposures ($\approx$\,1930~s) corresponds to 2.6\% of the orbital cycle. This means that the radial velocities of the UV Psc components can change, due to their orbital motion, by $\sim$\,20 km\,s$^{-1}$ during a single Stokes $V$ observing sequence. The effect of this radial velocity variation is that signals from the surface will be smeared out in the observed circular polarisation profiles, which needs to be taken into account in the data analysis. 

\section{Least-squares deconvolution profiles}
\label{sec:LSD}
Just like for most other cool active stars, the circular polarisation signatures of UV Psc are not strong enough to be reliably detected in individual spectral lines. In order to meaningfully study the magnetic field topology of the components of UV Psc, a multi-line polarisation diagnostic technique is required.

This study utilises the least-squares deconvolution technique \citep[LSD,][]{donati:1997}, as implemented by \cite{kochukhov:2010}, in order to obtain high-quality mean line profiles for both the Stokes $I$ and $V$ spectra. The construction of LSD profiles is built around approximating the stellar spectr\cla{um} as a superposition of copies of \cla{a} mean profile, scaled by the line depth and shifted to the central wavelength for each line considered. By finding the LSD profile that most accurately represents the spectr\cla{um} a significant improvement in S/N can be achieved. 

The scaling and shifting factors are combined into an LSD line mask that contains information about all lines in the observed wavelength regions. The line information was obtained from the VALD3 database \citep{ryabchikova:2015}, using MARCS \citep{gustafsson:2008} model atmospheres with effective temperatures closest to the observed effective temperatures, $T_{\mathrm{eff}} = 5780\pm100$ and $4750\pm80$~K, and surface gravity, $\log g=4.340\pm0.018$ and $4.478\pm0.019$, from \cite{popper:1997} for UV Psc A and B respectively. \cla{The spectral contributions of UV Psc components are not separated sufficiently in velocity for the multi-component LSD \citep{tkachenko:2013} to be viable. Consequently, a single line mask had to be adopted. We ended up using the mask corresponding to the parameters of the primary $T_{\mathrm{eff}} = 5750$~K and $\log g=4.5$. A cooler mask optimised for the secondary did not yield any advantages in detecting its polarisation signatures.} Some spectral regions in the observed wavelength range were excluded due to telluric absorption or broad stellar spectral features \cla{(e.g. hydrogen Balmer lines, Na D doublet, Ca HK and infrared triplet lines)}. The lines were then selected provided their residual depth was greater than 0.2 of the continuum. There were 3596 lines satisfying this criterion. The LSD line weights were normalised according to the mean wavelength $\lambda_{0}=5286$~\AA\ and effective Land\'e factor $\bar{g}_{0} = 1.21$.

\begin{figure}
    \centering
    \includegraphics[width=\hsize]{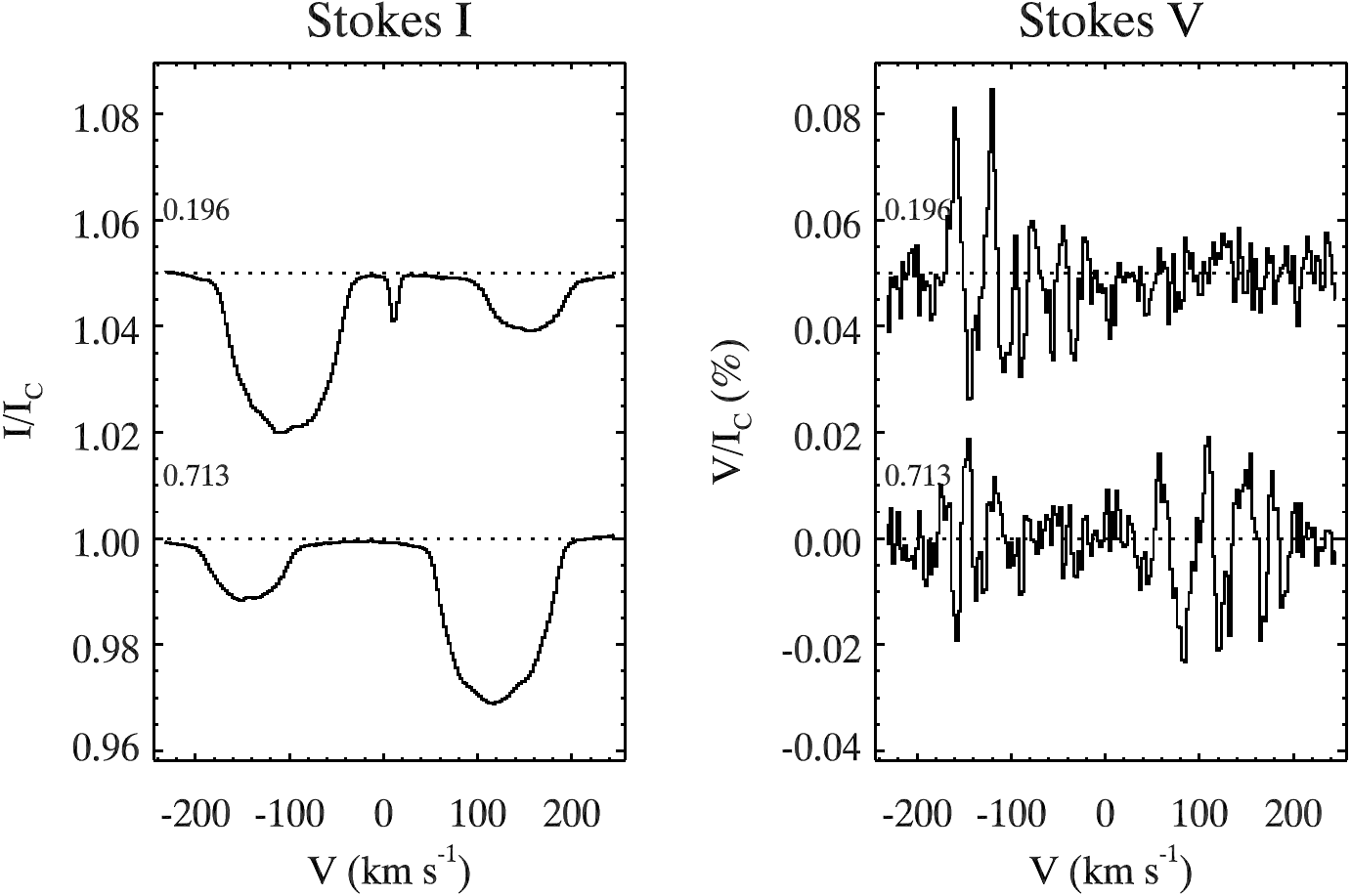}
    \caption{Stokes $I$ and $V$ LSD profiles of UV Psc for two of the observed phases. Stokes $V$ is shown as per cent of Stokes $I$ continuum. The profiles are shifted vertically for visibility. The contamination by the sunlight reflected off the Moon can be seen at phase 0.196.}
    \label{fig:LSDprofiles}
\end{figure}

The LSD technique was applied to each observation resulting in a total of 18 average Stokes $V$  line profiles. For Stokes $I$ the technique was applied separately to each sub-exposure, providing a total of 72 average line profiles. Example LSD profiles can be seen in Fig. \ref{fig:LSDprofiles}. It was discovered that four observation sequences of UV Psc obtained on September 17 were acquired when the object was close in the sky to the full Moon. This resulted in contamination by the reflected solar spectrum that was responsible for a narrow component visible in the Stokes $I$ profiles for this observing night. The solar contribution had to be removed before the ZDI analysis could be carried out. This was accomplished by fitting the affected intensity profiles with three Gaussians, two of which representing components of UV Psc that were broadened by the rotation rates of the respective star. The third Gaussian component represented the solar contamination and was not broadened by any rotation.

The obtained Stokes $V$ LSD profiles have an average S/N of 15950, which corresponds to a gain of about 50 compared to \cla{the} observed Stokes $V$ spectra. S/N of individual LSD profiles can be seen in Table~\ref{tab:StokesV}. With this gain, the LSD profiles can be used to assess if magnetic fields are detected on the star. This is done by considering the probability that the signal is compatible with zero \citep{donati:1992}. This is known as the false alarm probability (FAP) statistical analysis. In this study, a magnetic field detection is considered definite if the FAP is smaller than $10^{-5}$. A marginal detection is when the FAP is between $10^{-3}$ and $10^{-5}$. These thresholds were introduced by \cite{donati:1997} and are widely used within stellar spectropolarimetry research. For UV Psc A definite detections were achieved for all Stokes $V$ profiles but one, which still corresponded to a marginal detection. The global magnetic field of UV Psc B proved more challenging to detect with most LSD profiles. Only one marginal and one definite detection was made. This definite detection does however correspond to a phase when the profiles from the two components are overlapping. The polarisation signature of UV Psc A likely affects this result. The marginal detection corresponded to the phase 0.713 (see Fig. \ref{fig:LSDprofiles}) and had a FAP value ($1.5\times10^{-5}$) close to the definite detection threshold. \cla{No signatures corresponding to either component were detected in the LSD profiles computed from the diagnostic null spectra.}

\section{Spectral disentangling and orbital solution}
\label{sec:SpecOrb}
Since UV Psc is a spectroscopic binary, it becomes significantly more challenging to perform analysis of the individual components of the system compared to a single star. The spectrum of each component of the system needs to be separated from the contribution of the other star in order to study them quantitatively. In spite of complex and time-dependent line blending of the two components, it is possible to take advantage of multiple observations to extract information about the radial velocities and spectra of individual components. This procedure is known as spectral disentangling.

The first step is to obtain the radial velocity measurements for each component and observation. To accomplish this, the 72 Stokes $I$ LSD spectra were analysed using the disentangling code described in \cite{folsom:2010} and used in other studies of spectroscopic binaries \citep{rosen:2018,kochukhov:2019,lavail:2020}. This procedure iteratively performs least-squares fitting in order to derive the radial velocity shifts as well as mean profiles. A key assumption of this method is that the composite spectra can be described by the superposition of two shifted constant profiles, each originating from the individual binary component. This makes it challenging for the method to obtain accurate radial velocity estimates during phases when the two components are eclipsing each other. For this reason two observations had to be discarded when performing this analysis. On the other hand, the rotational line profile variability of UV Psc components due to temperature spots is weak enough to not significantly disturb the radial velocity measurements obtained with this technique.

\begin{figure}
   \centering
   \includegraphics[width=\hsize]{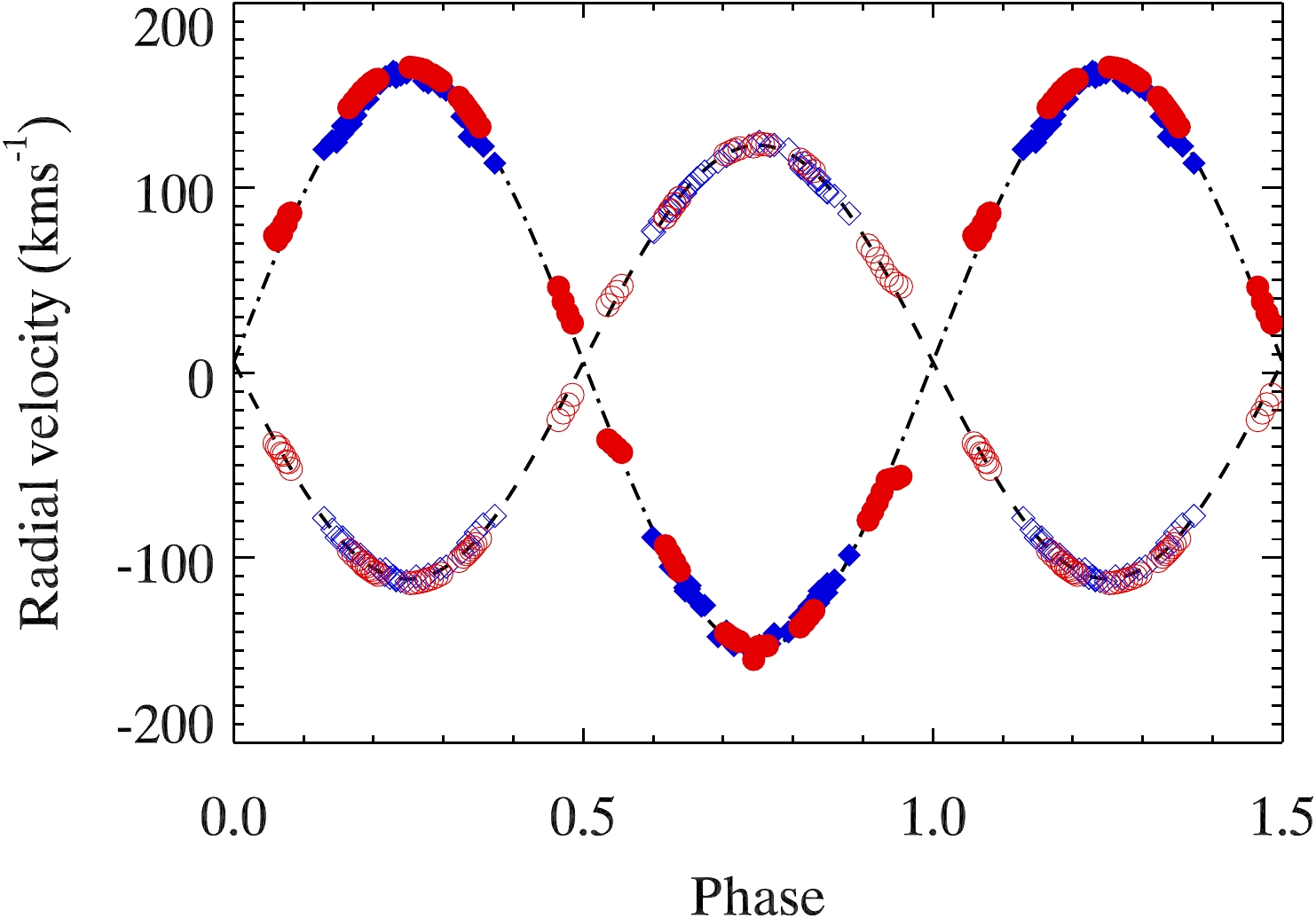}
      \caption{Orbital solution obtained for the two components of UV Psc. The circles (red) correspond to our observations with ESPaDOnS in 2016, the diamonds (blue) are measurements by \cite{popper:1997}. Filled symbols represent radial velocities for UV Psc B. The best-fitting orbits for the primary and secondary are indicated with the dashed and dash-dotted lines, respectively.}
         \label{fig:orbit}
   \end{figure}

The resulting radial velocities for the components of UV Psc are provided in Table~\ref{tab:RV}. The radial velocity error is determined by calculating the standard deviation between the orbital fit and observations. The resulting error is 1.73 and 3.77 km\,s$^{-1}$ for UV Psc A and B, respectively. From these data and previous radial velocity measurements by \citet{popper:1997}, the orbital parameters of the system were determined with the help of the least-squares fitting code \citep{tokovinin:1992}. The final orbital fits compared to the observed radial velocities are shown in Fig.~\ref{fig:orbit}. The corresponding orbital parameters are reported in Table~\ref{tab:orbit}, where some were fitted while others were adopted from previous studies of UV Psc \citep{popper:1997,kjurkchieva:2005}. Based on the revised spectroscopic orbit and the estimate of the orbital inclination \cla{$i=88.9\pm2\degr$}, given by \citet{popper:1997} \cla{and confirmed by \citet{kjurkchieva:2005}}, we calculated masses  \cla{$M_{A}=1.0225\pm0.0058$} $M_\sun$ and \cla{$M_{B}=0.7741\pm0.0034$} $M_\sun$ for the primary and secondary respectively. \cla{These masses are in between the previous determinations by \citet[][$M_{A}=0.9829\pm0.0077$ $M_\sun$, $M_{B}=0.7644\pm0.0045$ $M_\sun$]{torres:2010} 
and \citet[][$M_{A}=1.10\pm0.03$ $M_\sun$, $M_{B}=0.81\pm0.03$ $M_\sun$]{kjurkchieva:2005}}.

\begin{table}
      \caption[]{
      Log of spectropolarimetric observations of UV Psc taken at CFHT.
      }
         \label{tab:StokesV}
         \begin{tabular}{llll}
            \hline \hline
            \noalign{\smallskip}
            Reduced HJD  &  Phase & S/N$_{V}$ & S/N$_{LSD}$\\
            \noalign{\smallskip}
            \hline
            \noalign{\smallskip}
	          57604.1136 & 0.753 & 322 & 17487\\
    	      57607.1346 & 0.261 & 252 & 13402\\
    	      57649.8900 & 0.917 & 229 & 11966\\
    	      57649.9134 & 0.944 & 246 & 13059\\
    	      57650.0200 & 0.068 & 310 & 16452\\
    	      57650.1303 & 0.196 & 321 & 17144\\
    	      57650.8840 & 0.071 & 320 & 17218\\
    	      57650.9725 & 0.174 & 324 & 17236\\
    	      57651.0693 & 0.286 & 320 & 17015\\
    	      57651.8502 & 0.193 & 315 & 16731\\
    	      57651.9781 & 0.342 & 301 & 15757\\
    	      57652.0919 & 0.474 & 280 & 14778\\
    	      57652.8302 & 0.332 & 309 & 16332\\
    	      57653.0843 & 0.544 & 292 & 15486\\
    	      57653.0135 & 0.627 & 302 & 15944\\
    	      57653.9458 & 0.627 & 322 & 17069\\
    	      57654.0196 & 0.713 & 323 & 17151\\
    	      57654.1113 & 0.819 & 318 & 16873\\
            \noalign{\smallskip}
            \hline
         \end{tabular}
   \end{table}

\begin{table}
  \caption[]{Orbital solution for UV Psc.}
  \label{tab:orbit}
  \begin{tabular}{ll}
    \hline\hline
    \noalign{\smallskip}
    Parameter & Value\\
    \hline
    \multicolumn{2}{c}{Fitted quantities:} \\
    $P_{\mathrm{orb}}$ (d) & 0.86104716$^\mathrm{a}$ \\
    HJD$_{0}$ & $2448897.4226\pm0.0003$\\
    $K_{A}$ (km\,s$^{-1}$) & $117.20\pm0.18$\\
    $K_{B}$ (km\,s$^{-1}$) & $154.81\pm0.37$\\
    $\gamma$ (km\,s$^{-1}$) & $5.79\pm0.13$\\
    $e$ & $0.0^{\mathrm{b}}$\\
    \hline
    \multicolumn{2}{c}{Derived quantities:} \\
    $M_{A}$ ($M_{\sun}$) & \cla{$1.0225\pm0.0058^{\mathrm{c}}$}\\
    $M_{B}$ ($M_{\sun}$) & \cla{$0.7741\pm0.0034^{\mathrm{c}}$}\\
    \hline
  \end{tabular}
  \tablefoot{(a) Adopted from \cite{kjurkchieva:2005}. (b) Circular orbit was assumed. (c) Calculated using $i=88.9\pm2\degr$ from \cite{popper:1997}.
  }
\end{table}

\begin{figure}
\centering
\includegraphics[width=\hsize]{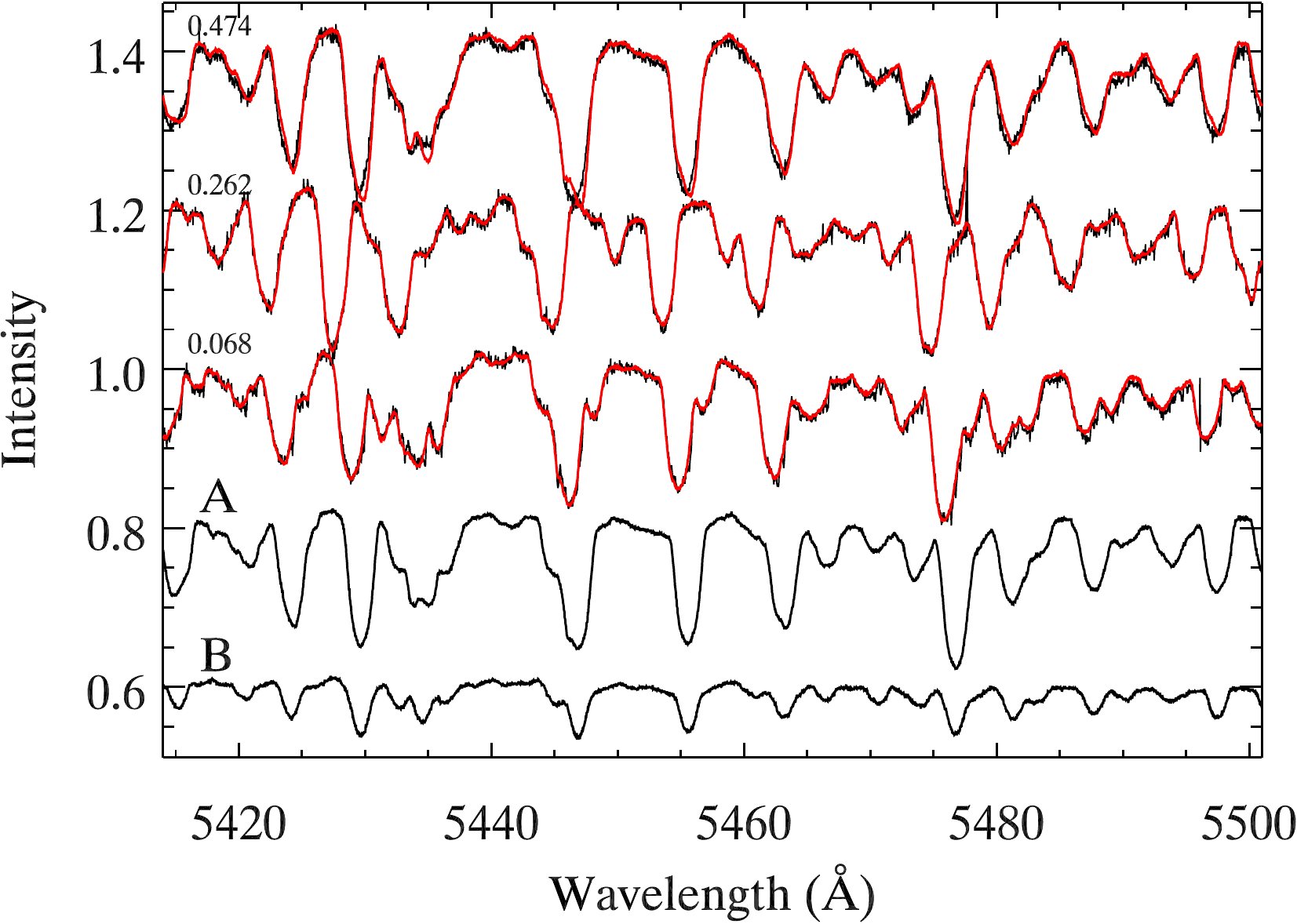}
\caption{Illustration of the spectral disentangling analysis. The two bottom curves show the separated spectra for each component. These two spectra are combined into composite spectra that are compared to observations. The two top curves show this comparison for two representative orbital phases\cla{, one near the eclipse and one at furthest separation}. For visibility the spectra have been shifted vertically.}
\label{fig:disentangle}
\end{figure}

In order to perform a Zeeman intensification analysis, another disentangling procedure must be applied. This is because individual lines are required for estimating magnetic field strength with this method. Spectral disentangling was carried out around lines most suitable for a Zeeman intensification study. Considering solar-like properties of UV Psc A, we used the lines from the \ion{Fe}{i} multiplet employed in the analysis of active solar analogues \citep{kochukhov:2020a}, which is known to contain some of the most magnetically sensitive lines in the optical spectra of Sun-like stars. In order to obtain the spectra of the individual components of UV Psc, the disentangling was carried out by representing the observed spectra as superposition\cla{s} of two stellar components. The radial velocity shifts were adopted according to the orbital solution obtained above. The model was then fitted to all Stokes $I$ observations to yield disentangled, time-averaged spectra of UV Psc A and B. An illustration of this procedure is presented in Fig.~\ref{fig:disentangle}, which shows the obtained \cla{mean spectra for each component} in the wavelength region containing some of the magnetically sensitive \ion{Fe}{i} lines and the fit \cla{of the mean spectra to two observations at different orbital phases. Here the mean spectra for each component have been shifted according to its radial velocity at the given phase}.

While the disentangling procedure is capable of separating the line contributions of each component of UV Psc, it does not account for the continuum dilution. To obtain spectra that can be compared to synthetic models, the disentangled spectra of each component must be re-scaled using the luminosity ratio of the system \citep[e.g.][]{folsom:2008}. Previous analyses of UV Psc \citep{kjurkchieva:2005} found a luminosity ratio ranging between $\sim$ 5.0 and 7.5 in the V band. An intermediate value of 6 is used as \cla{an} initial guess for re-scaling of the disentangled spectra in the context of our Zeeman intensification analysis.

\section{Zeeman-Doppler imaging}
\label{sec:ZDI}
The DI and ZDI inversion techniques are powerful methods to obtain information on surface structures that are not directly observable by other means. In this work, the InversLSDB binary ZDI code described by \cite{rosen:2018} were used to obtain both the surface brightness distribution and magnetic field topology. This code is a development of the InversLSD code \citep{kochukhov:2014}. It enables mapping of multiple components in a binary system simultaneously using observed LSD intensity and polarisation profiles. The spectral modelling implemented in InversLSDB accounts for eclipses and can be performed by either assuming spherical stellar shapes and arbitrary orbits, or by using the Roche lobe geometry to describe corotating binary components with aligned orbital and rotational axes. For close binaries, such as UV Psc, the second approach based on the Roche lobe geometry is suitable and were adopted for our study. In order to carry out the analysis, the code needs a set of stellar parameters for each component. These include the mass of each component, orbital period and inclination, and the two values of Roche surface potentials. \cla{The latter were adjusted to yield the equivalent volume radii equal to the measured radii, $R_A=1.110\pm0.023$~$R_\odot$ and $R_B=0.835\pm0.018$~$R_\odot$ \citep{torres:2010}, of the components. The Roche lobe treatment resulted in a slight deviation from spherical geometry. The difference between the polar radii and the radii pointing towards the other component were $\sim$ 3\% and 2\% for UV Psc A and B, respectively. The specific values can be seen in Table \ref{tab:rocheRadius}.} The code also relies on free parameters that can not be determined from the prior knowledge of the system. The relative local brightness between the two components needs to be fixed before ZDI can be carried out. \cla{The relative local brightness indicates the difference in brightness between the surface elements of equal area on the two components. 
} This parameter was determined \cla{to be $I_A/I_B=2.09$} by finding the best fit to the Stokes $I$ spectra, assuming a uniform surface temperature \cla{distribution}.

\begin{table}
    \caption{\cla{Radii of the components of UV Psc. $R_{\mathrm{vol}}$ is the equivalent volume radius, $R_{\mathrm{in}}$ is the radius pointing towards the other component, $R_{\mathrm{side}}$ is the equatorial radius perpendicular to $R_{\mathrm{in}}$, and $R_{\mathrm{pol}}$ is the polar radius.}}
    \label{tab:rocheRadius}
    \begin{tabular}{lcc}
        \hline \hline
        \cla{Radii ($R_\sun$)} & \cla{A} & \cla{B} \\
        \hline
        \cla{$R_{\mathrm{vol}}$ } & \cla{1.111} & \cla{0.836} \\
        \cla{$R_{\mathrm{in}}$ }& \cla{1.134} & \cla{0.847} \\
        \cla{$R_{\mathrm{side}}$} & \cla{1.109} & \cla{0.834} \\
        \cla{$R_{\mathrm{pol}}$ }& \cla{1.096} & \cla{0.829} \\
        \hline
    \end{tabular}
\end{table}

During the reconstruction process, it was found that the LSD profiles obtained for two phases could not be reproduced by the model. Further investigation showed that these two phases correspond to the two observations made in August 2016, before the rest of the data set. This could imply a change of the magnetic field structure in the span of around a month. For this reason, these two observations were excluded from further analysis of the magnetic field structure of the components of UV Psc.

\subsection{Brightness distribution}
\label{ch:DI}
\begin{figure*}
    \centering
    \includegraphics[width=\textwidth]{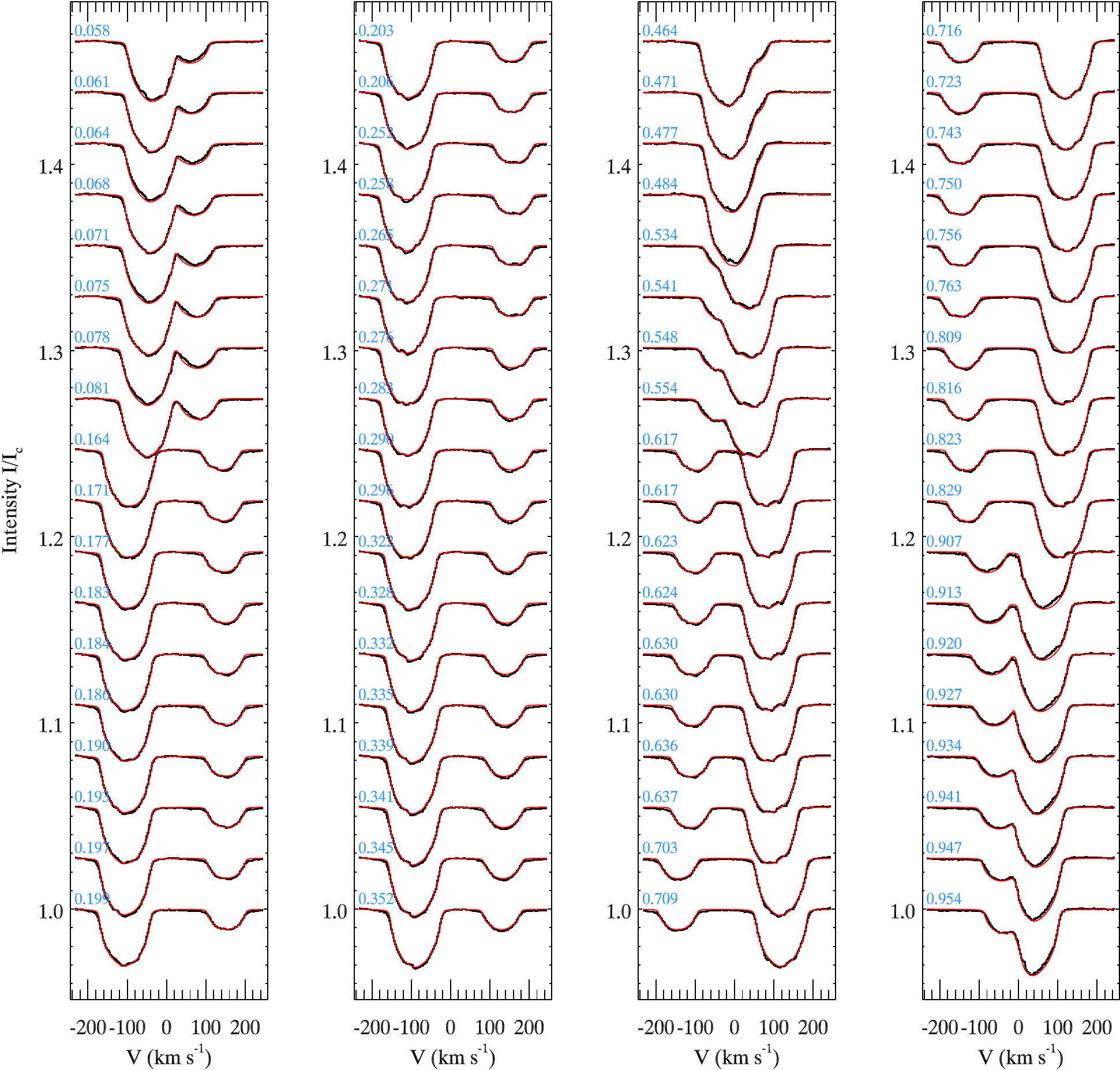}
    \caption{Comparison of the observed (black histogram lines) and model (red solid lines) Stokes $I$ LSD profiles corresponding to brightness reconstruction. The phase of each observation is indicated on the left.
    The profiles are shifted vertically for visibility.
    }
    \label{fig:stokesI}
\end{figure*}

\begin{figure*}
\centering
\includegraphics[width = 0.70\textwidth]{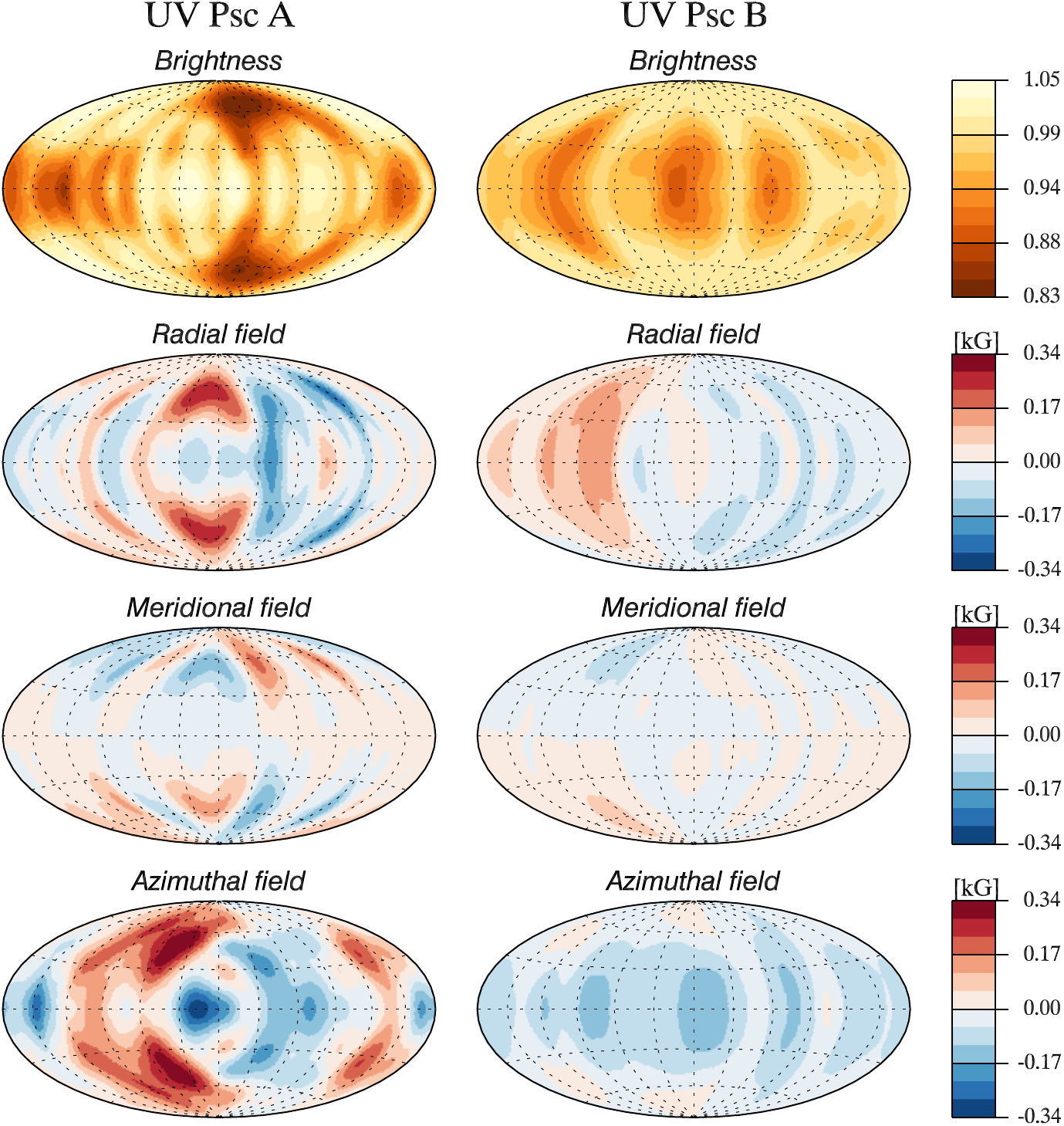}
\includegraphics[width = 0.25\textwidth]{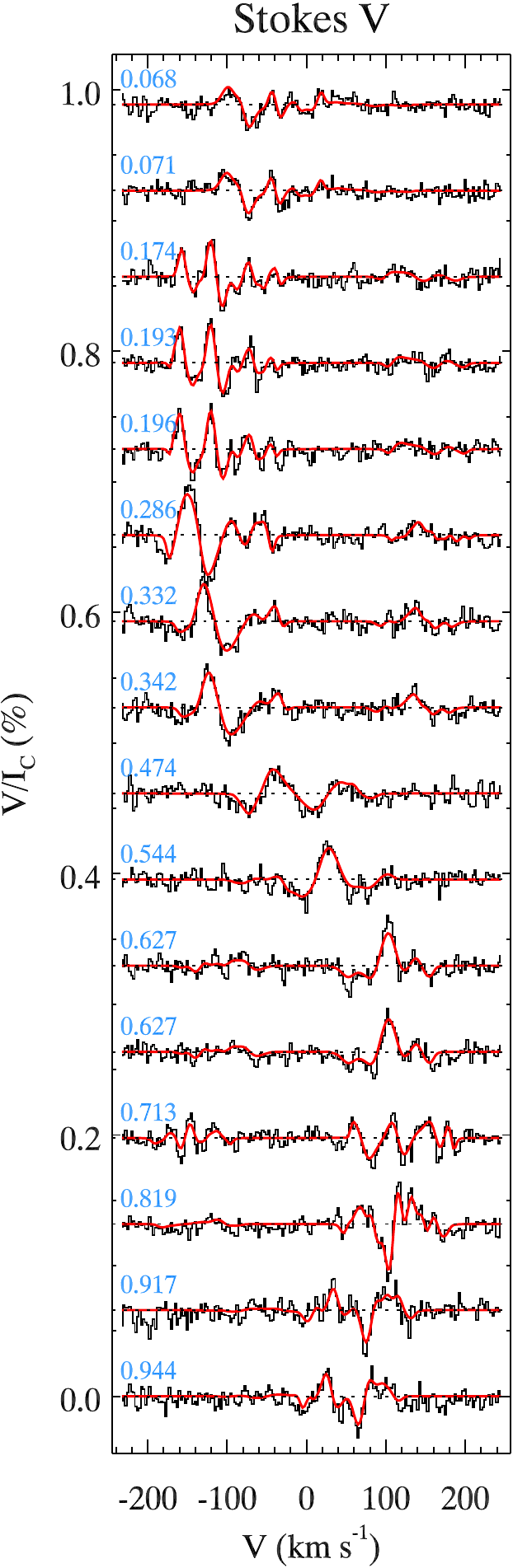}
\caption{\textit{Left}. Maps of the brightness distribution (top row), obtained from the Stokes $I$ spectra and the magnetic field components (row 2--4), obtained from the Stokes $V$ spectra for UV Psc A (left column) and UV Psc B (right column). The surface is represented using the Hammer-Aitoff projection in order to preserve the relative area of different latitude bands. The central meridian corresponds to a longitude of $180\degr$. \cla{The substellar points are located at the longitude $0\degr$ for UV Psc A and at $180\degr$ for UV Psc B.} The colour bars on the right indicate the relative surface brightness and magnetic field strength in kG for each map. \textit{Right}. Observed (black histogram) and best-fit spectra (red solid lines) obtained from the ZDI analysis. The spectra for different phases are offset vertically, with the phase indicated next to each spectrum.} 
\label{fig:map}
\end{figure*}

The brightness distribution on the stellar surface is determined by applying DI to the 72 individual Stokes $I$ LSD profiles. While accounting for a non-uniform brightness distribution is important in order to not underestimate field intensity, \cite{rosen:2012} showed that, as long as the field is \cla{weaker than a few kG}, temperature inhomogeneities dominate \cla{the} variability of the Stokes $I$ profiles. This means that the brightness distribution can be reconstructed independently of the magnetic field in the case of UV Psc.

The best fit Stokes $I$ profiles are shown in Fig.~\ref{fig:stokesI}. These profiles correspond to the final brightness maps derived for the components of UV Psc, shown in Fig.~\ref{fig:map}. 
The maps were obtained with Tikhonov regularisation, following the procedure described by \cite{rosen:2018}. This introduces two regularisation parameters\cla{, $\Lambda_{T}^{(i)}$,} in addition to the $\chi^2$-function in order to favour certain solutions. Specifically, this regularisation tries to minimise the local brightness contrast according to the penalty function \cla{$R_{T}$},
\begin{equation}
R_{T}^{(1)}=\Lambda_{T}^{(1)}\sum_{i}\sum_{j}(T_{i}-T_{j(i)})^2.
\end{equation}
The first sum is carried over \cla{the surface brightness $T$ of} every surface element on each component of UV Psc while the second sum is carried out over the $j$ adjacent elements to the $i$th element. Since DI does not restrict the average brightness of the star, this requires the introduction of a secondary regularisation function that favours small deviation from the default surface brightness value ($T_0=1$),
\begin{equation}
R_{T}^{(2)}=\Lambda_{T}^{(2)}\begin{cases}
\sum_{i}C_{1}(T_{i}-T_{0})^2,\indent\text{if } T_{i}\leq T_{0}\\
\sum_{i}C_{2}(T_{i}-T_{0})^2,\indent\text{if } T_{i}>T_{0}\\
\end{cases}.
\end{equation}
This regularisation also introduces the opportunity to penalise dark or hot spots to a different degree by changing the $C_{1}$ or $C_{2}$ parameters, respectively. In the case of this analysis of UV Psc the values were set to $C_{1}=1$ and $C_{2}=10$, following the approach of \cite{rosen:2018}. \cla{The stronger suppression of bright spots is justified by the general observation that dark, cool features dominate the surface structure of active late-type stars \citep[e.g.][]{berdyugina:2005}.
}

Due to the fact that UV Psc is an eclipsing binary, its inclination angle is close to $90^{\circ}$. This makes DI incapable of distinguishing between the northern and southern hemisphere of the components. This is why the obtained surface brightness distributions are symmetric with respect to the equators.

\subsection{Magnetic field topology}
\label{sec:Bfield}
The ZDI technique was applied to the Stokes $V$ LSD profiles in order to obtain the surface magnetic field topologies for the two components of UV Psc. Information about the surface brightness distributions obtained from the procedure in Sect.~\ref{ch:DI} is also accounted for by InversLSDB in the magnetic geometry reconstruction. It is important to be aware of the fact that the ZDI \cla{ reconstruction does not provide complete information on the stellar magnetic fields.} The method is only sensitive to large- and intermediate-scale magnetic fields while signals from small-scale fields are largely cancelled by other nearby areas with opposite magnetic field polarities. For a more complete picture these small-scale magnetic fields also need to be studied using Zeeman intensification (discussed later in Sect.~\ref{ch:ZIntensification}).

The magnetic field was reconstructed, following the inversion procedure introduced by \cite{kochukhov:2014} and implemented for binary stars by \cite{rosen:2018} and \citet{kochukhov:2019}. This includes a spherical harmonic description of the magnetic field with three sets of spherical harmonic coefficients, $\alpha_{\ell,m}$, $\beta_{\ell,m}$, and $\gamma_{\ell,m}$, describing the radial poloidal, horizontal poloidal, and toroidal field components respectively. These coefficients are penalised with a regularisation function that favours weaker and simpler fields corresponding to smaller values of $\alpha$, $\beta$, $\gamma$, and lower angular order $\ell$,
\begin{equation}
R_{B} = \Lambda_{B}\sum_{\ell=1}^{\ell_{\mathrm{max}}}\sum_{m=-\ell}^{\ell}\ell^2(\alpha_{\ell,m}^2+\beta_{\ell,m}^2+\gamma_{\ell,m}^2).
\end{equation}
Here $\ell_{\mathrm{max}}$ is the maximum angular degree that is accounted for in the calculations, it can be selected either based on the smallest scale potentially visible according to the \cla{instrumental profile of the spectropolarimeter} and $v_{\rm e}\sin i$ of the star or as a value beyond which no contribution to the reconstructed map can be seen.  For this study we selected an $\ell_{\mathrm{max}}$ of 30. \cla{This choice was based on the observation that harmonics at this and higher orders made no contribution to the magnetic field, even at weaker regularisation than what was adopted for the reconstruction.}

The treatment of local synthetic LSD profiles is based on the Unno-Rachkovsky formulae \citep{landi-deglinnocenti:2004}. This assumes that the LSD profiles respond to \cla{the} magnetic field in a way equivalent to a Zeeman triplet with an effective Land\'e factor $\bar{g}=1.21$ and wavelength $\lambda=5286$~\AA, both corresponding to \cla{the} mean parameters of the LSD mask. As mentioned in Sect.~\ref{ch:obs}, the polarisation observations with ESPaDOnS are constructed from multiple subexposures that resulted in an observation time \cla{that was} not negligible compared to the orbital period of the components of UV Psc. Since details of surface structures are of interest for this study, phase smearing must be considered in order to accurately reconstruct the surface magnetic fields. This was implemented by calculating five separate profiles over the phase interval covered by each observation and then integrating over this interval.

The obtained magnetic field maps can be seen in Fig.~\ref{fig:map}. The map shows stronger fields on UV Psc A compared to UV Psc B, even if UV Psc B is less luminous this is in agreement with fewer definite detections from the FAP calculations in Sect.~\ref{sec:LSD}. The strongest field component of the primary is the azimuthal component, with the weakest contribution coming from the meridional magnetic field. This is a common occurrence in ZDI inversions when no linear polarisation is included. It has been both observed \citep{rosen:2015} and demonstrated numerically \citep{kochukhov:2002,rosen:2012} that without linear polarisation, a crosstalk between the radial and meridional components is introduced. This causes the meridional field components to be systematically underestimated.

To quantify \cla{the} magnetic properties of the stars in \cla{the} UV Psc system, we used the average magnetic field strength, determined from,
\begin{equation}
\langle B \rangle = \frac{\sum_{i}S^{i}\sqrt{\left.B_{r}^{i}\right.^2 + \left.B_{m}^{i}\right.^2 + \left.B_{a}^{i}\right.^2}}{\sum_{i}S^{i}}.
\end{equation}
Where $S^{i}$ are surface element areas for each star. The average magnetic fields are 137~G and 88~G for UV Psc A and UV Psc B respectively. The maximum magnetic field strengths obtained on the surface of UV Psc A and B are 429 and 202\,G, respectively. In order to compare global field characteristics of different stars, the distribution of the magnetic energy is commonly used. Table \ref{tab:energyDist} shows the contributions of the poloidal, toroidal, and axisymmetric (according to two different definitions) magnetic field to the total magnetic energy in per cent. Here the axis of symmetry refers to the rotational axis. These values show that the field geometry of UV Psc A is dominated by a poloidal field that is not symmetric with respect to the rotational axis. UV Psc B has a more even distribution between poloidal and toroidal fields. It also has about 50\% of the energy contained in \cla{the} axisymmetric magnetic component. 

\begin{figure}[!t]
\centering
\includegraphics[width=0.5\textwidth]{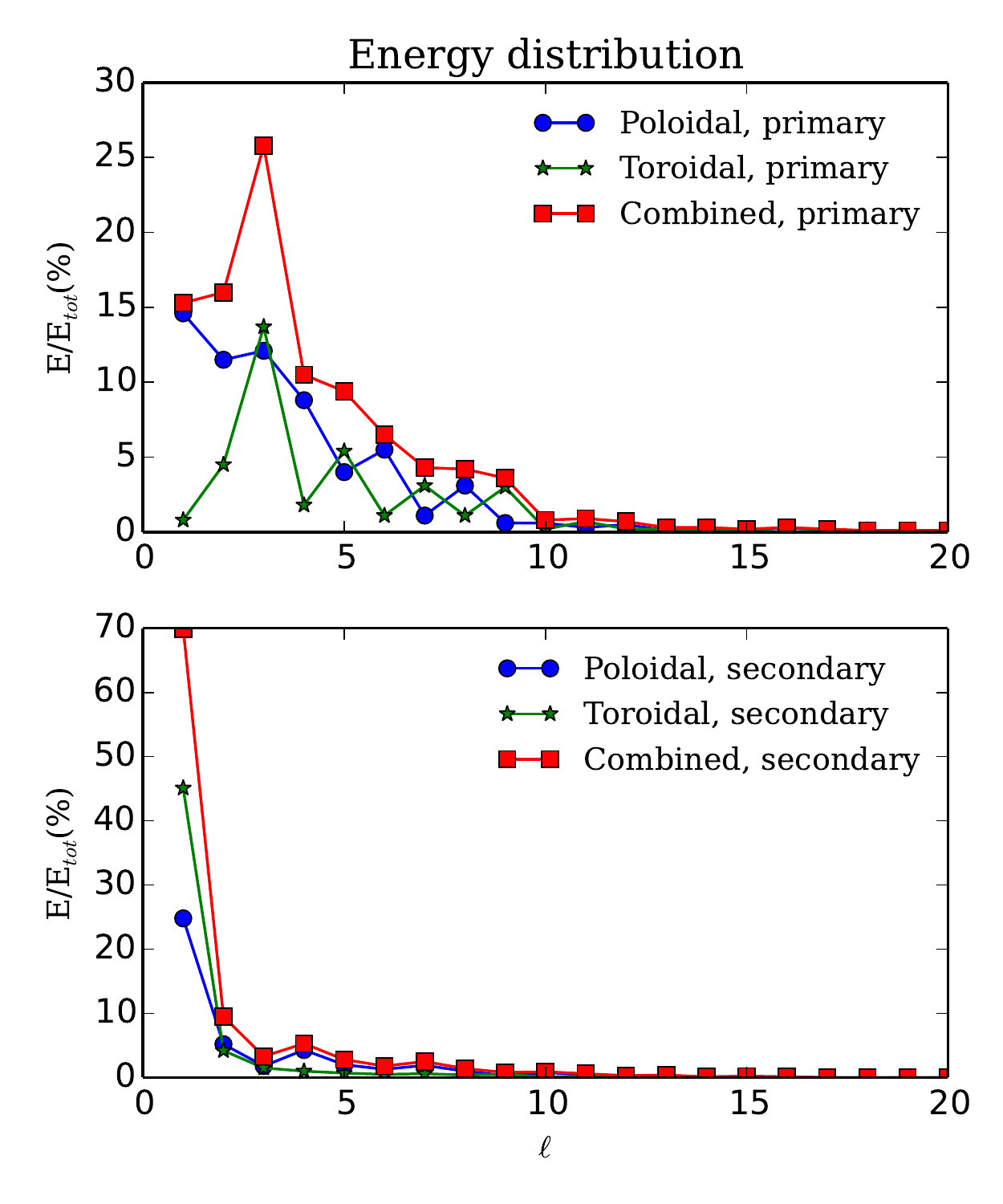}
\caption{Relative magnetic energy as a function of the spherical harmonic angular degree $\ell$. \cla{\textit{Top.} Distribution of the poloidal, toroidal, and combined magnetic field energy over different angular degrees for UV Psc A. \textit{Bottom.} Same as above but for UV Psc B.}}
\label{fig:EnergyDist}
\end{figure}

How the energy is distributed among the different spherical harmonic modes also provides a useful information \cla{with which} to compare \cla{the} global fields of different stars. The distribution of magnetic energy, as a function of the angular degree $\ell$ for both components of UV Psc can be seen in Fig.~\ref{fig:EnergyDist}. The field of UV Psc B is dominated by low order modes, with around 70\% contained in the lowest order $\ell=1$. Contributions of higher modes quickly decreases to a few \%. For UV Psc A the situation is different, no angular order dominate to the same extent as for UV Psc B. The largest contribution comes from the angular order $\ell=3$. For both components the contributions from the angular orders beyond $\ell\approx10$ is almost negligible. One cause for this could be the lack of linear polarisation, which has been shown \citep{rosen:2015} to cause higher order modes to be underestimated. The weaker signal originating from UV Psc B could further enhance this effect, 
resulting in the inversion procedure being able to reconstruct only the lowest order modes.

\begin{table}
  \caption[]{Fraction of magnetic energy contained in the poloidal and axisymmetric harmonic field components.}
  \label{tab:energyDist}
  \begin{tabular}{lcc}
    \hline \hline
    Component & Primary & Secondary\\
    \hline
    $E_{\mathrm{pol}}/E_{\mathrm{tot}}$(\%) & 64.0 & 44.9 \\
    $E_{m=0}/E_{\mathrm{tot}}$(\%) & 15.7 & 49.1 \\ $E_{|m|<\ell/2}/E_{\mathrm{tot}}$ (\%) & 35.0 & 52.4\\
    \hline
  \end{tabular}
\end{table}

\subsection{Hemisphere degeneracy}
\label{sec:degeneracy}
The result from both surface brightness and magnetic field reconstruction shows surface features that are symmetric with respect to the equator. This comes from the method's difficulty in disentangling signals originating from each hemisphere. This is a well-known phenomenon that has been observed and discussed in previous studies of eclipsing binaries and other DI targets with a high inclination angle \citep{vincent:1993,kochukhov:2019}. The issues resulting in this hemisphere degeneracy are however different for the surface brightness and magnetic field maps.

\subsubsection{Effect on the brightness distribution}

\begin{figure}[!t]
  \centering
  \includegraphics[width=0.5\textwidth]{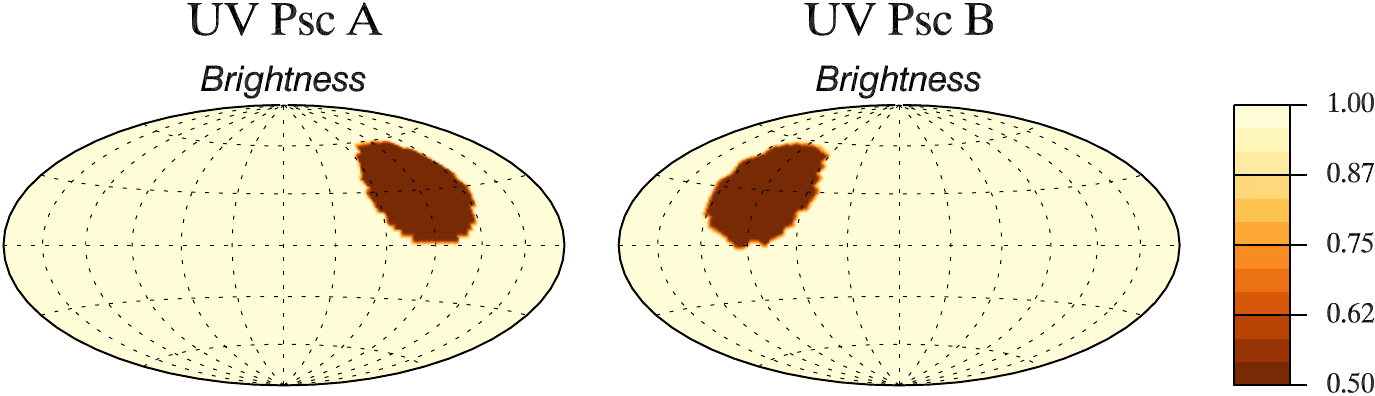}
  \includegraphics[width=0.5\textwidth]{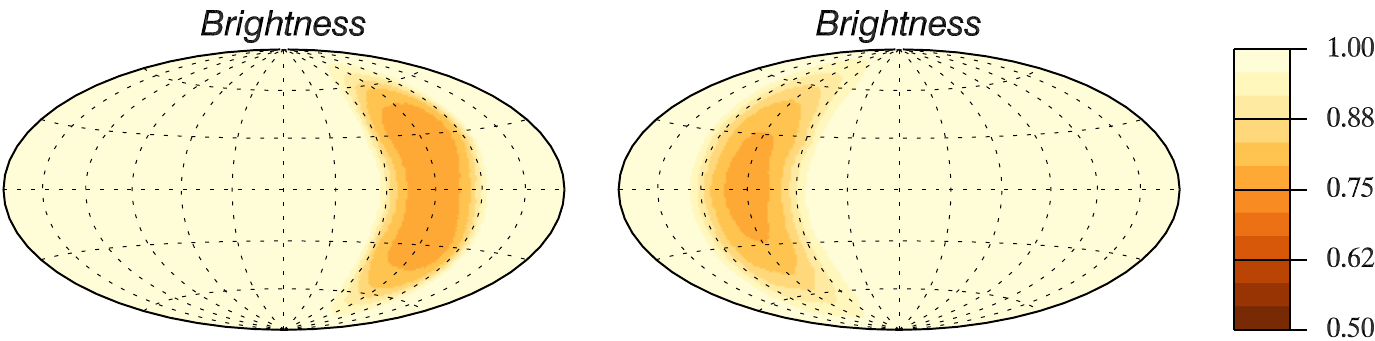}
  \caption{\textit{Top}. Synthetic surface spots generated on each component. \textit{Bottom}. Result obtained after fitting the Stokes $I$ spectra from the synthetic brightness distribution using binary-star DI.}
  \label{fig:synthSpot}
\end{figure}

In order to assess the effects causing the degeneracy, we moved away from the complex observations and considered a simple synthetic surface brightness map comprising a single dark spot on the surface of each component, as seen at the top of Fig.~\ref{fig:synthSpot}. The geometry of the system was in this case identical to UV Psc. From this surface structure, synthetic spectra were generated and then used to reconstruct the surface in the same way as was done for the real observations. This reconstruction can be seen below the original spots in Fig.~\ref{fig:synthSpot}. What can be seen is that the contrast of dark spots has been greatly reduced and the spots have been smeared out over a large range of latitudes, both above and below the stellar equators, while retaining their longitudinal extension. This smearing occurs due to the fact that the spectral contributions from surface elements on each side of the equator are not distinguishable. In other systems, where the inclination is not close to 90$^\circ$, the limb darkening breaks this degeneracy. This allows DI to determine which hemisphere a surface feature belongs to. 

One consequence of the hemisphere degeneracy in brightness mapping is that this could lead to a misinterpretation of the Stokes $V$ profiles used to determine the magnetic field structure. \cite{rosen:2012} have shown that reconstruction\cla{s} of magnetic fields are dependent on accurate brightness maps. 
However, in the case of UV Psc the surface brightness features are smeared in latitude and their contrast is underestimated. This could result in further errors in the magnetic field estimation.

\subsubsection{Effect on the magnetic field geometry}
\begin{figure*}
  \centering
  \includegraphics[width=0.48\textwidth]{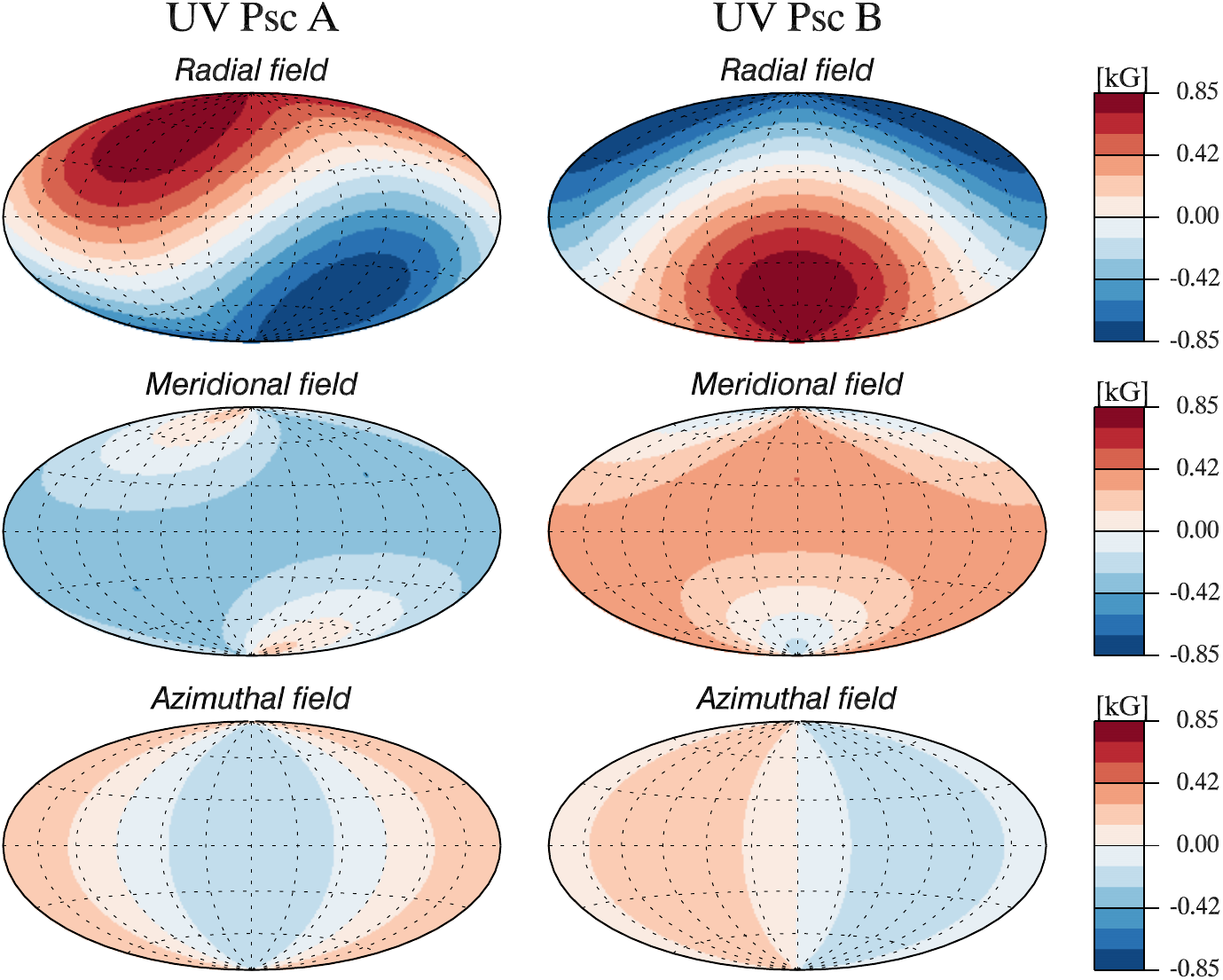}
  \includegraphics[width=0.48\textwidth]{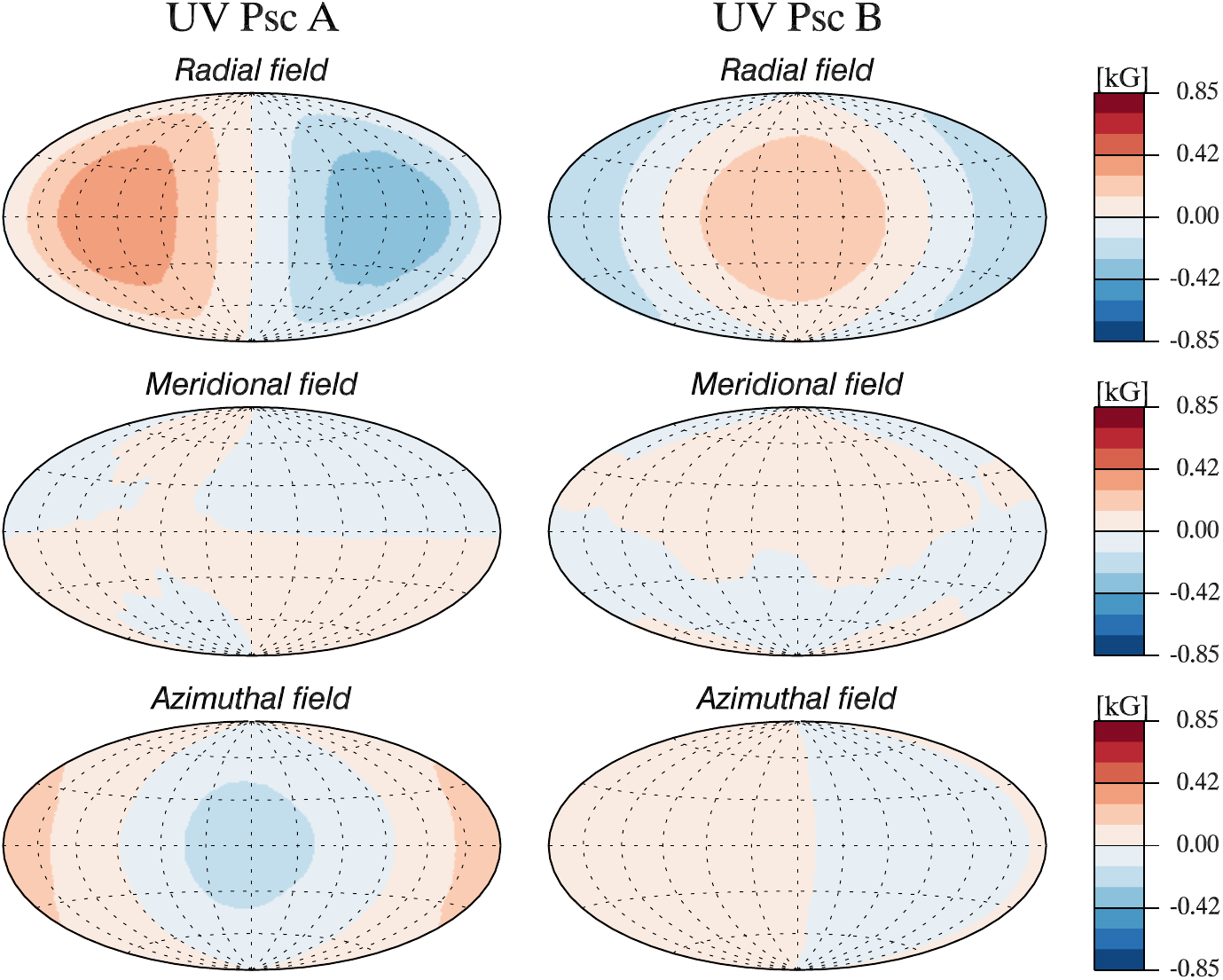}
  \caption{\textit{Left}. 
Inclined dipolar fields on each component of UV Psc employed for generation of simulated observations.
\textit{Right}. Magnetic structure reconstructed from simulated observations using the same ZDI inversion procedure that was used to obtain the maps in Fig.~\ref{fig:map}.
}
  \label{fig:synB}
\end{figure*}
\begin{figure*}
\centering
\includegraphics[height=0.6\textheight]{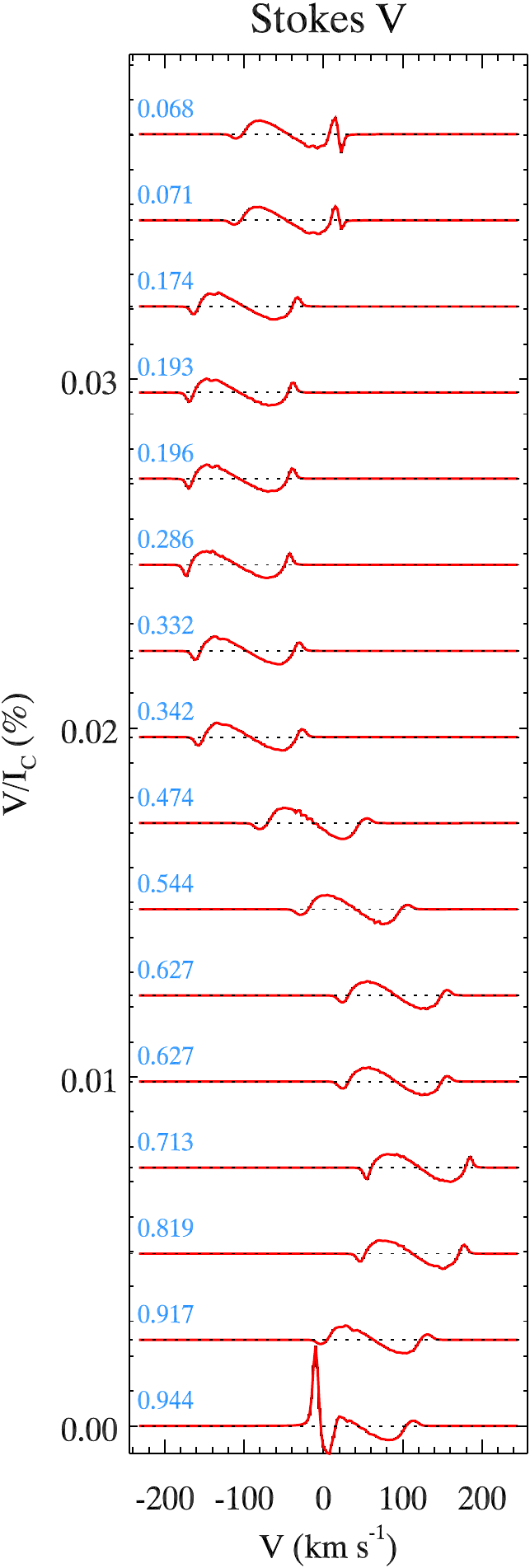}
\includegraphics[height=0.6\textheight]{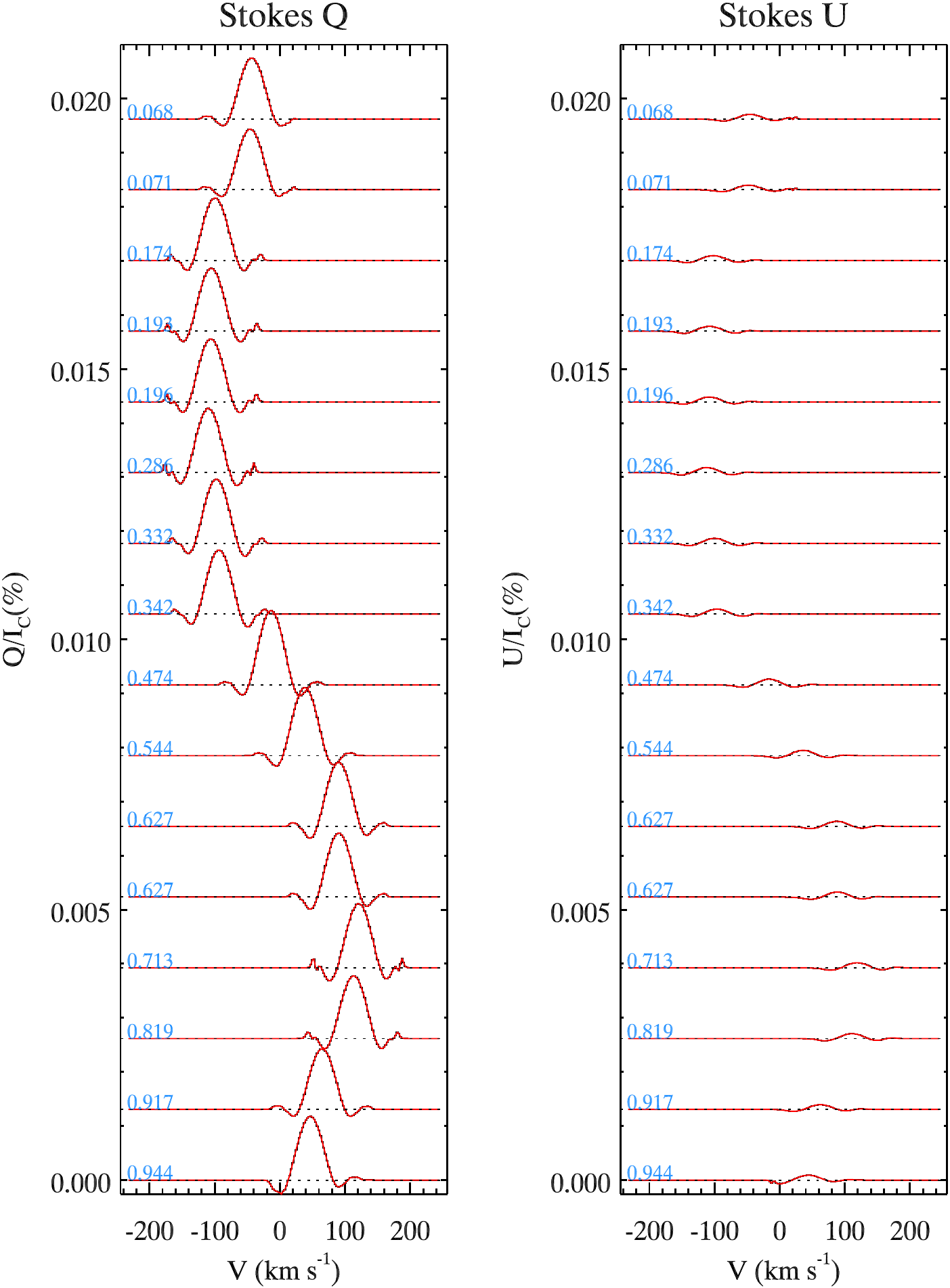}
\caption{Polarisation profiles corresponding to a dipole, antisymmetric with respect to the equator. All profiles are shifted vertically for visibility. The profile strengths are given in per cent relative to continuum of Stokes $I$ profiles.}
\label{fig:Testprofiles}
\end{figure*}

To obtain a better idea of what causes the symmetric structures in the magnetic field maps, a simpler synthetic dipolar field was constructed on the surface of both components of UV Psc. The initial structure can be seen in Fig.~\ref{fig:synB}. These field geometries corresponds to 1~kG dipolar fields with different signs, inclined at 45$^\circ$ with respect to the stellar rotational axes.
Generating synthetic spectra, adding realistic random noise, and applying the same ZDI technique as for the actual observational data yields reconstructed field distributions shown in Fig.~\ref{fig:synB}. This demonstrates that the method only reproduces the part of the dipole that is symmetric relative to the equator while the antisymmetric component entirely disappears. To further investigate this, magnetic field structures corresponding to specific spherical harmonic coefficients were constructed. This was done for each mode with $\ell\leq2$, one example corresponding to $\ell=1$, $m=0$ can be seen in Fig.~\ref{fig:Testprofiles}. In this antisymmetric configuration the amplitude of the Stokes $V$ profiles is significantly lower compared to our observations. A signal increase can be seen for the phase 0.944, corresponding to a temporary break of symmetry when the secondary eclipses the primary. Similar results were found for other antisymmetric harmonic modes. This shows that the magnetic field components that originate from spherical harmonic modes that are antisymmetric with respect to the equator, are mostly cancelled out on a large scale as well as on a small scale. This is another reason why the global magnetic field strengths of eclipsing binaries are likely to be underestimated by ZDI studies.

Another effect of the magnetic inversion from simulated data is that the magnetic field on the smaller component UV Psc B is more underestimated than the magnetic field on UV Psc A. Since the secondary is both colder and smaller than the primary its spectral contribution is inevitably weaker and likely more suppressed by the noise of the observations.

\cla{It is interesting to consider how different inclinations affect the reconstruction bias discussed above. To do this, inversions for inclination angles of 85$^\circ$ and 80$^\circ$ were also simulated keeping other parameters the same. A simple numerical comparison can be made by considering how accurately the magnetic energy of symmetric and anti-symmetric modes was recovered. The values for the different inclinations can be seen in Table~\ref{tab:energyReconstruction}. There is an improvement in recovering anti-symmetric structures as inclination is reduced, although the reconstructed field is still dominated by the symmetric component. The corresponding magnetic field maps are shown in Fig.~\ref{fig:SynBInclination}.}

\begin{table}[]
\cla{
    \caption{Relative energies corresponding to the symmetric and anti-symmetric $\ell=1$ harmonic modes reconstructed for the primary at different inclinations. 
    }
    \label{tab:energyReconstruction}
    \begin{tabular}{c|cc}
        \hline \hline
        Inclination & Symmetric & Anti-symmetric \\
        \hline
        Input & 50\% & 50\% \\
        88.9$^\circ$ & 93.4\% & 0.4\% \\
        85.0$^\circ$ & 76.7\% & 12.5\%\\
        80.0$^\circ$ & 74.2\% & 16.3\% \\
        \hline
    \end{tabular}
}
\end{table}

Since the observations analysed in our study have only been carried out with circular polarisation it is interesting to investigate if the hemisphere degeneracy can be broken with the introduction of linear polarisation. Keeping the stellar geometry the same, we calculated linear polarisation profiles corresponding to a dipole on UV Psc A that is aligned with the rotational axis and antisymmetric with respect to the equator. The resulting Stokes $QU$ profiles take the form seen in Fig.~\ref{fig:Testprofiles}. Some signal in the linear polarisation is visible, even for the spherical harmonic modes where the contribution from the circular polarisation is undetectable. The maximum strength of linear polarisation is in this case larger than circular polarisation, except during the eclipse caused by the secondary component. This means that linear polarisation profiles could be used to obtain more information on the magnetic field structures of stars with an inclination close to $90^{\circ}$. \cla{However, obtaining time-resolved linear polarisation observation for UV Psc with necessary $S/N$ ratio is not feasible with ESPaDOnS and will likely require spectropolarimetric observations at 8-m class telescopes \citep[e.g.][]{strassmeier:2018}.}

\section{Zeeman intensification}
\label{ch:ZIntensification}
As mentioned in the previous section, the ZDI technique only probes large- and intermediate-scale magnetic field structures. In order to get a complete picture of the magnetic field on the stellar surface, the small-scale fields need to be studied as well. As found by many studies \citep[e.g.][]{see:2019,kochukhov:2020a}, most of the magnetic energy of a cool star is contained within these small scales. This means that the magnetic structures neglected by ZDI are very important for determining the overall intensity of the field and its effect on the star. 

For UV Psc the mean magnetic field strength was determined using the Zeeman intensification method, \cla{applied to the mean disentangled spectra of the components}. This approach follows the work by \citet{kochukhov:2020a} and \citet{kochukhov:2020b}. In these studies groups of spectral lines with different sensitivity to the Zeeman effect were used to obtain information on the average small-scale magnetic field on different stars. One difficulty with studying the Zeeman intensification on UV Psc is that the components are close together and tidally locked, resulting in high $v_{\rm e}\sin{i}$ for both components. As a consequence, many of the lines used in previous Zeeman intensification studies of Sun-like stars are blended, requiring other lines to be selected. Here we attempted to use the same multiplet of \ion{Fe}{i} lines as was employed by \citet{kochukhov:2020a}. Two unblended lines could be found, \ion{Fe}{i} 5429.7 and \ion{Fe}{i} 5497.5~\AA. The latter line has been used previously and was identified to be the \ion{Fe}{i} line with the largest relative strength change due to a magnetic field in the optical spectrum of a Sun-like star. The former line has a relatively weak response to the field, thus providing a suitable reference for determination of non-magnetic parameters.

Similar to \citet{kochukhov:2020a}, our Zeeman intensification modelling of UV Psc components relied on calculating theoretical spectra using the polarised radiative transfer code Synmast \citep{kochukhov:2010}. The line parameters were taken from VALD and MARCS model atmospheres were used. The stellar atmospheric parameters were the same as adopted in Sect.~\ref{sec:LSD}. The multi-component field strength distribution was parameterised according to \citet{lavail:2019}, with each component assumed to have a radial field with the strength increasing in 2~kG steps. The abundance of Fe was also varied in the synthetic spectral grid, in steps of 0.05 dex. The abundances of other elements were assumed to be equal to solar values according to \cite{asplund:2009}. A microturbulent velocity of 1~km\,s$^{-1}$ was assumed. The disentangled spectra, to which these model spectra were fitted, were produced as described in Sect.~\ref{sec:SpecOrb}. 

\cla{The} magnetic field is not the only parameter that affects stellar spectr\cla{a}, which results in some degeneracy between the magnetic field strength and other parameters. To explore these degeneracies and derive realistic uncertainties, the free parameters were optimised by implementing Markov Chain Monte-Carlo (MCMC) sampling using SoBAT routines \citep{anfinogentov:2020} written in IDL. To give the walker time to reach the highest likelihood region of the parameter space a burn-in length of 20,000 was used. The sampling was then carried out until it reached convergence. This was determined by calculating the effective sample size \citep[ESS,][]{sharma:2017}, using the autocorrelation time of the walker. Convergence was set to require the ESS to be larger than 1000.

We started with the analysis of the primary.
The parameters used in the sampling were the filling factors ($f_{i}$), each associated with non-zero magnetic field strengths. These filling factors were given uniform priors in the range [0,1] with the additional constraint, 
\begin{equation}
\sum_{i}f_{i} \leq 1,
\end{equation}
to ensure a physically sensible solution. The Fe abundance was also included as a free parameter. A uniform prior in the range $[-4.50,-4.70]$ was used for Fe abundance expressed in $\log (N_{\rm Fe}/N_{\rm tot})$ units. The projected rotational velocity $v_{\rm e}\sin{i}$ was given a uniform prior in the range [65,75] km\,s$^{-1}$\cla{. While this range extends beyond the rotation broadening expected from the radius and period, in the context of this inference the $v_{\rm e}\sin{i}$ parameter represents all contributions to non-magnetic broadening of spectral lines besides instrumental broadening. This includes, for example, the macroturbulent velocity broadening that was not explicitly accounted for in our analysis.} The radial velocity was given a uniform prior in the $[-2,2]$ km\,s$^{-1}$ range. A parameter to scale the continuum around each line was also included in the parameter inference, each was given a uniform prior [0.95,1.05]. The number of magnetic filling factors was varied during different runs. Each run was then compared to find the most suitable number of filling factors using the Bayesian Information Criterion \citep[BIC,][]{sharma:2017},
\begin{equation}
  \label{eq:BIC}
  \mathrm{BIC} = -2\ln{p(Y|\hat{\theta})}+d\ln{n}.
\end{equation}
The first term corresponds to the maximum likelihood, \cla{where $Y$ is the observed data and $\hat{\theta}$ is the highest likelihood parameters}. The second term is given by the number of data points ($d$) and free parameters ($n$). \cla{The number of filling factors for which the minimum value of Eq.~(\ref{eq:BIC}) is obtained is chosen as the optimal description of the magnetic field.} For UV Psc A it was found that the optimal number of filling factors was three, corresponding to a superposition of the components with 0, 2 and 4\,kG field strengths. \cla{No benefit was found by adding a fourth filling factor corresponding to 6\,kG.}

\begin{figure}
  \centering
  \includegraphics[width=\hsize]{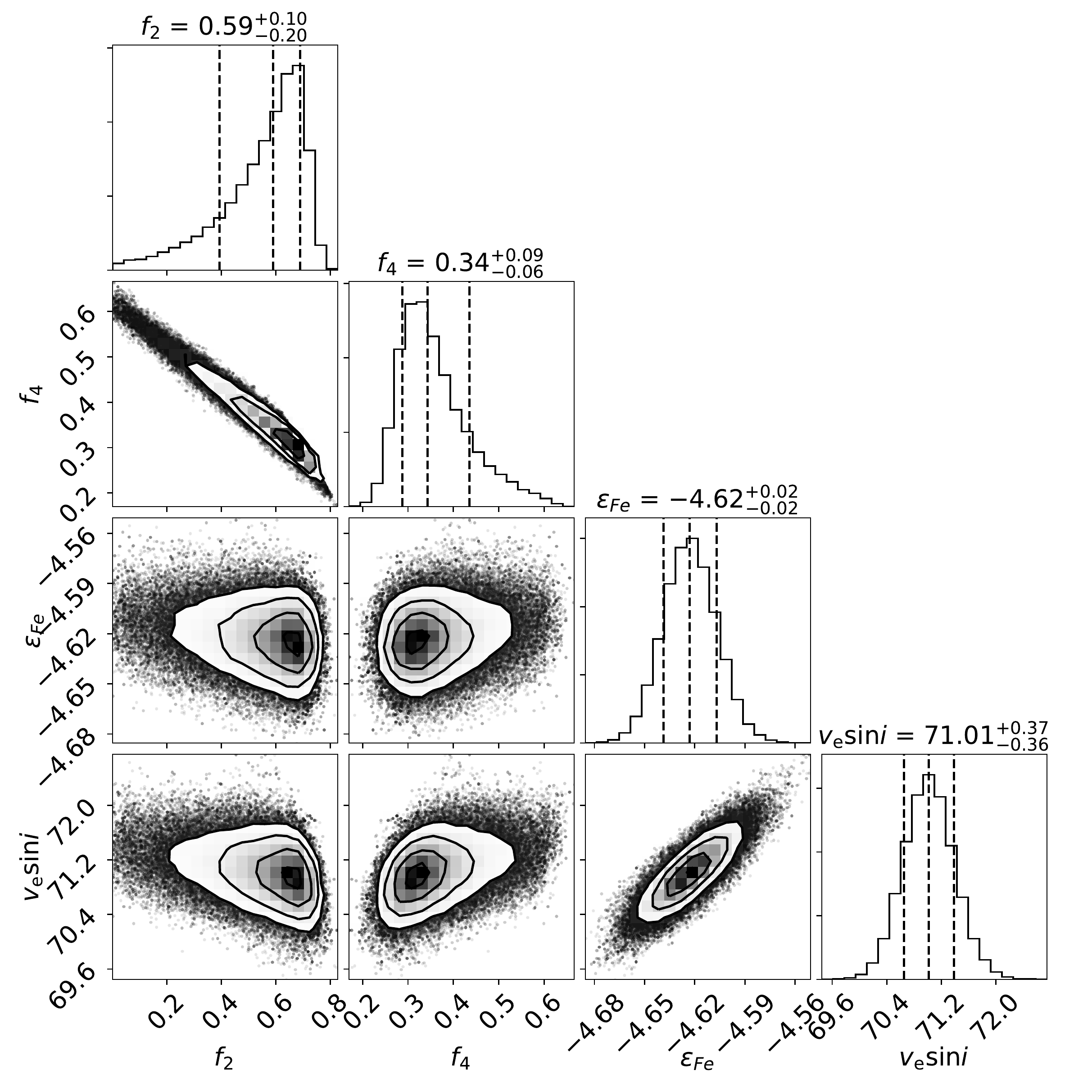}
  \includegraphics[width=\hsize]{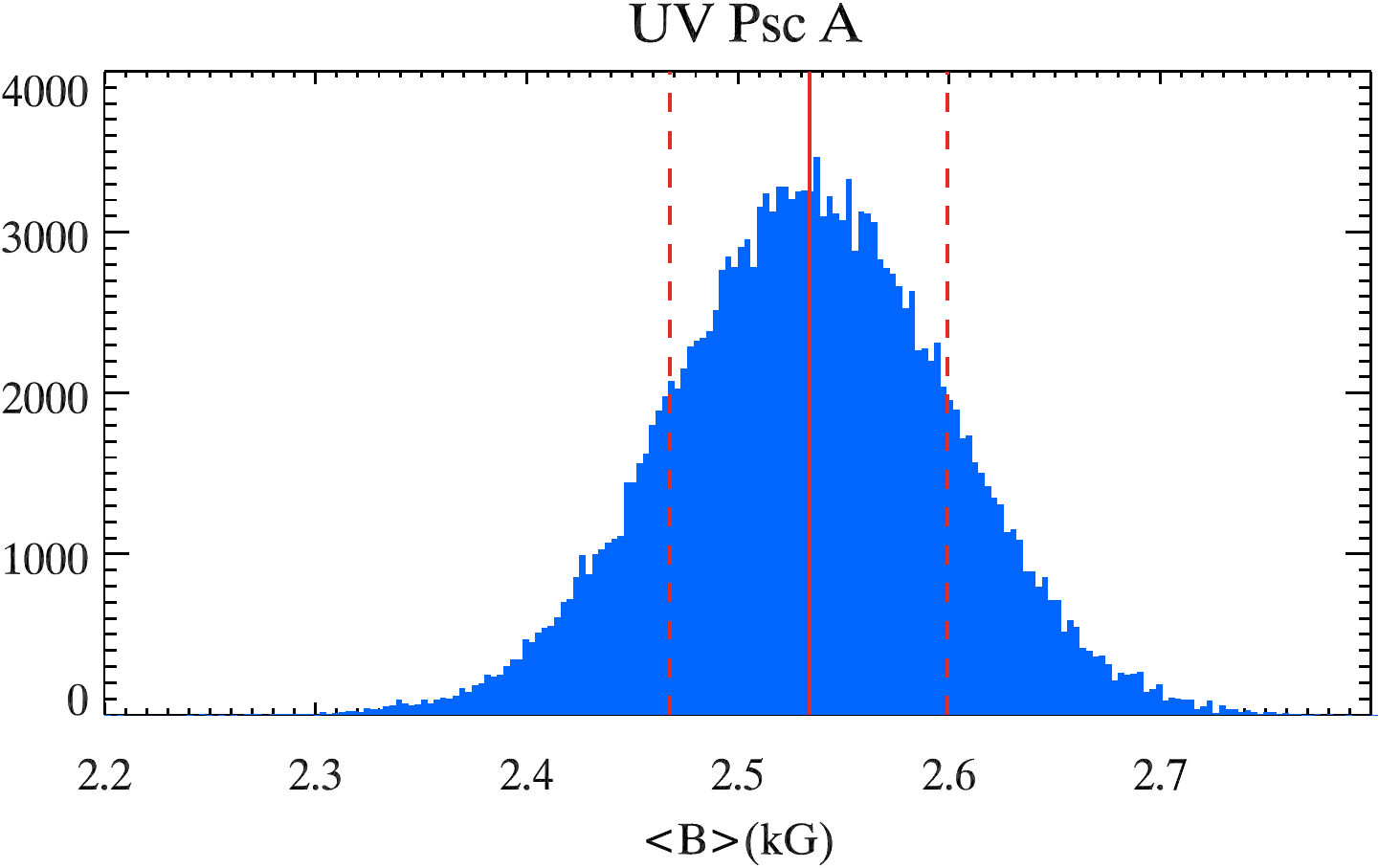}
  \caption{Illustration of the MCMC Zeeman intensification analysis of UV Psc A. \textit{Top}. Corner plot showing the posterior distribution of the model parameters (magnetic filling factors $f_{i}$, abundance $\varepsilon_{\mathrm{Fe}}$, rotational velocity $v_{\rm e}\sin{i}$). \textit{Bottom}. Posterior distribution of the resulting mean magnetic field strength plotted with its median value (red solid line) and the $1\sigma$ confidence region (red dashed lines).}
  \label{fig:cornerA}
\end{figure}

\begin{figure}
  \centering
  \includegraphics[width=\hsize]{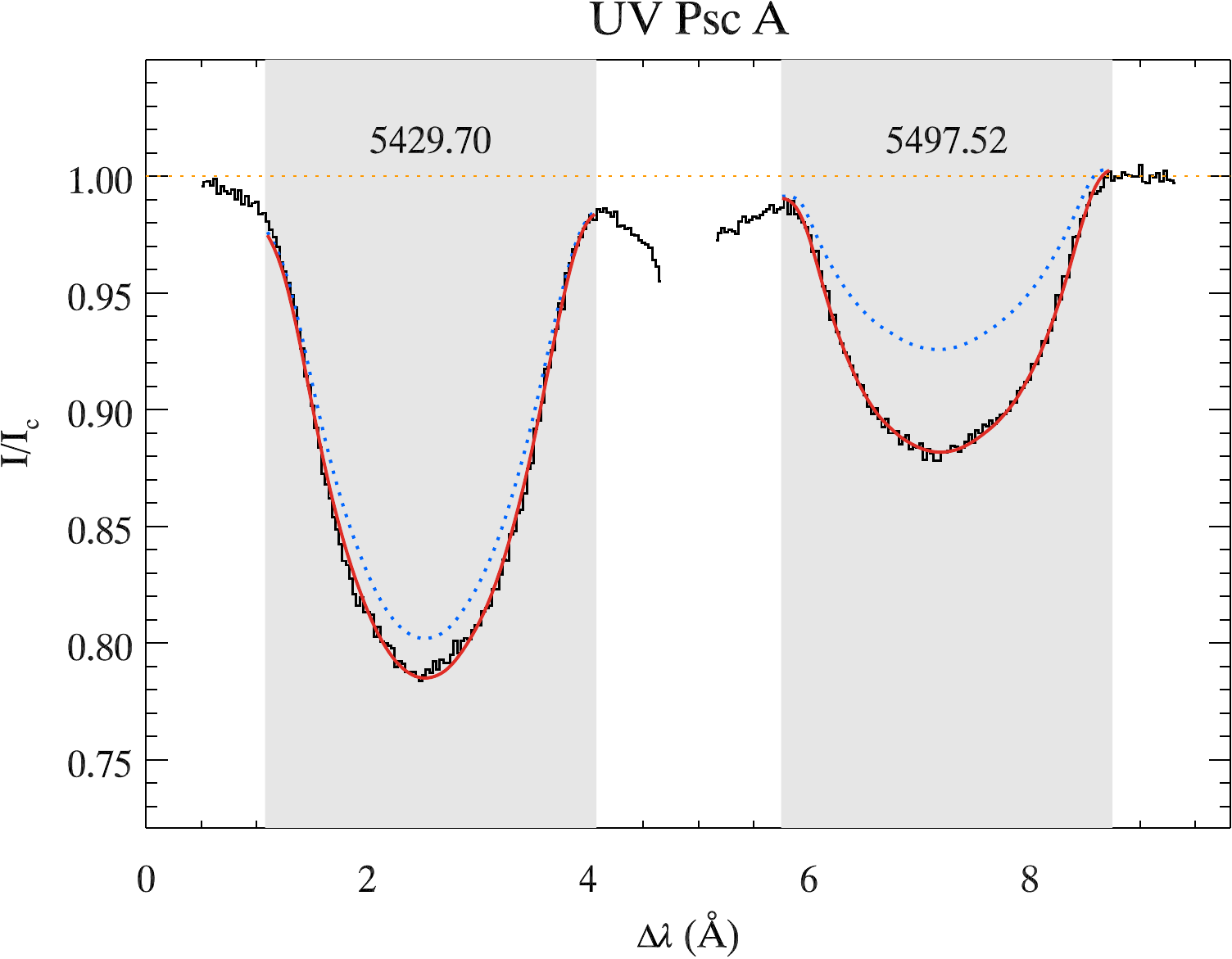}
  \caption{Profiles of the two \ion{Fe}{i} spectral lines employed for inferring the magnetic field strength for UV Psc A. Observations are shown in black histogram\cla{s} and are compared to the median parameter fit (red solid line). The blue dotted lines correspond to \cla{the} non-magnetic synthetic spectrum. The dark grey areas indicate the wavelength regions where the fitting has been carried out.}
  \label{fig:line_fit}
\end{figure}

The posterior parameter distributions corresponding to
the MCMC sampling for UV Psc A are shown in Fig.~\ref{fig:cornerA}. The median of the average magnetic field strength was found to be $\langle B\rangle=2.53\pm0.07$\,kG, where the uncertainty corresponds to $\pm1\sigma$ (68\% confidence) intervals. For the obtained median parameters, the model fit to the observed profiles of the two \ion{Fe}{i} lines is presented in Fig.~\ref{fig:line_fit}. For comparison, we also show non-magnetic theoretical spectra of the same two lines. There is a significant difference between the synthetic line profiles that include or exclude \cla{the} magnetic field, highlighting the need to use \cla{the} magnetic field in the analysis \cla{of the spectra} of UV Psc A spectra. From the sampling it is also clear that there is a strong degeneracy between the 2 and 4~kG magnetic components. This is likely the consequence of the high rotation rate of the star, which effectively smears out any Zeeman broadening due to the dominant Doppler broadening. This means that only the magnetic intensification of the \ion{Fe}{i} 5497.5~\AA\ \cla{line} is effectively providing information on the magnetic field. In this case, the effect of a weak field covering a large part of the star is essentially identical to that of a strong field covering a smaller part.

A similar analysis was then applied to the disentangled spectrum of UV Psc B. It is reasonable to assume the same element abundances for the two close binary components. Therefore, the Fe abundance is already constrained by the results obtained for UV Psc A. For this reason it was not included as a free parameter in the MCMC sampling. Instead, the line strength was adjusted with the help of the luminosity ratio which was treated as a free parameter. This is justified by the fact that previous empirical determinations of the luminosity ratio from light curves yielded a set of values spanning a significant range \citep{kjurkchieva:2005}. Since the secondary is less bright than the primary, the variation in the luminosity ratio will have a strong effect (similar to changing \cla{the} element abundance) on its spectral lines. The prior for the luminosity ratio was centred on $LR \equiv L_A/L_B=6$ with limits of $\pm2$ to cover the range of previous observations of the V band luminosity ratio. The prior for $v_{\rm e}\sin{i}$ was also adjusted to the range [45,55] km\,s$^{-1}$, in line with the expected equatorial rotation rate of UV Psc B. With these priors, the MCMC sampling was carried out with the same burn-in and ESS requirement as for UV Psc A. Similar to UV Psc A, the number of filling factors was varied and the resulting BIC values compared. This procedure also led to the final model with three filling factors corresponding to 0, 2 and 4~kG. 

\begin{figure}
  \centering
  \includegraphics[width=\hsize]{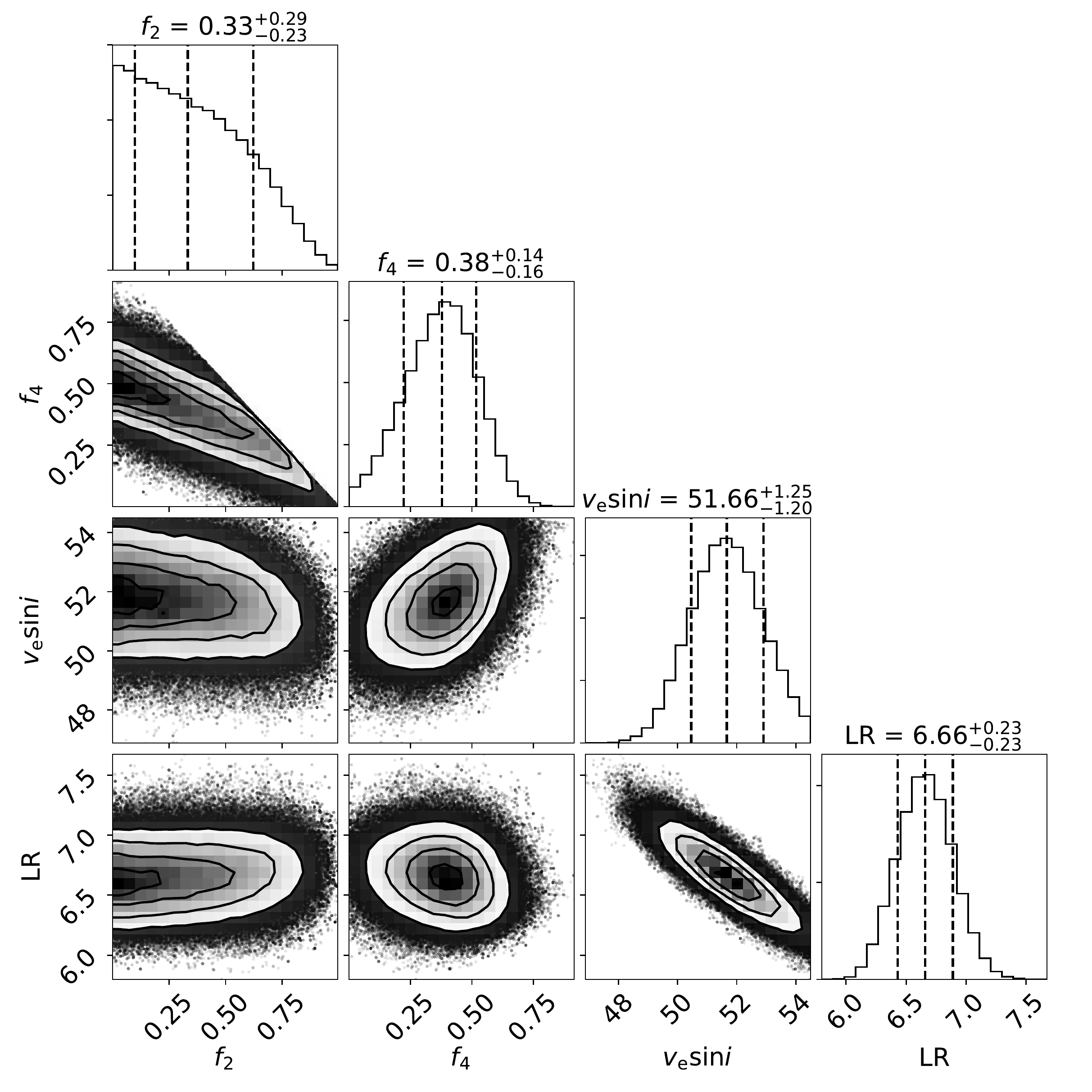}\\
  \includegraphics[width=\hsize]{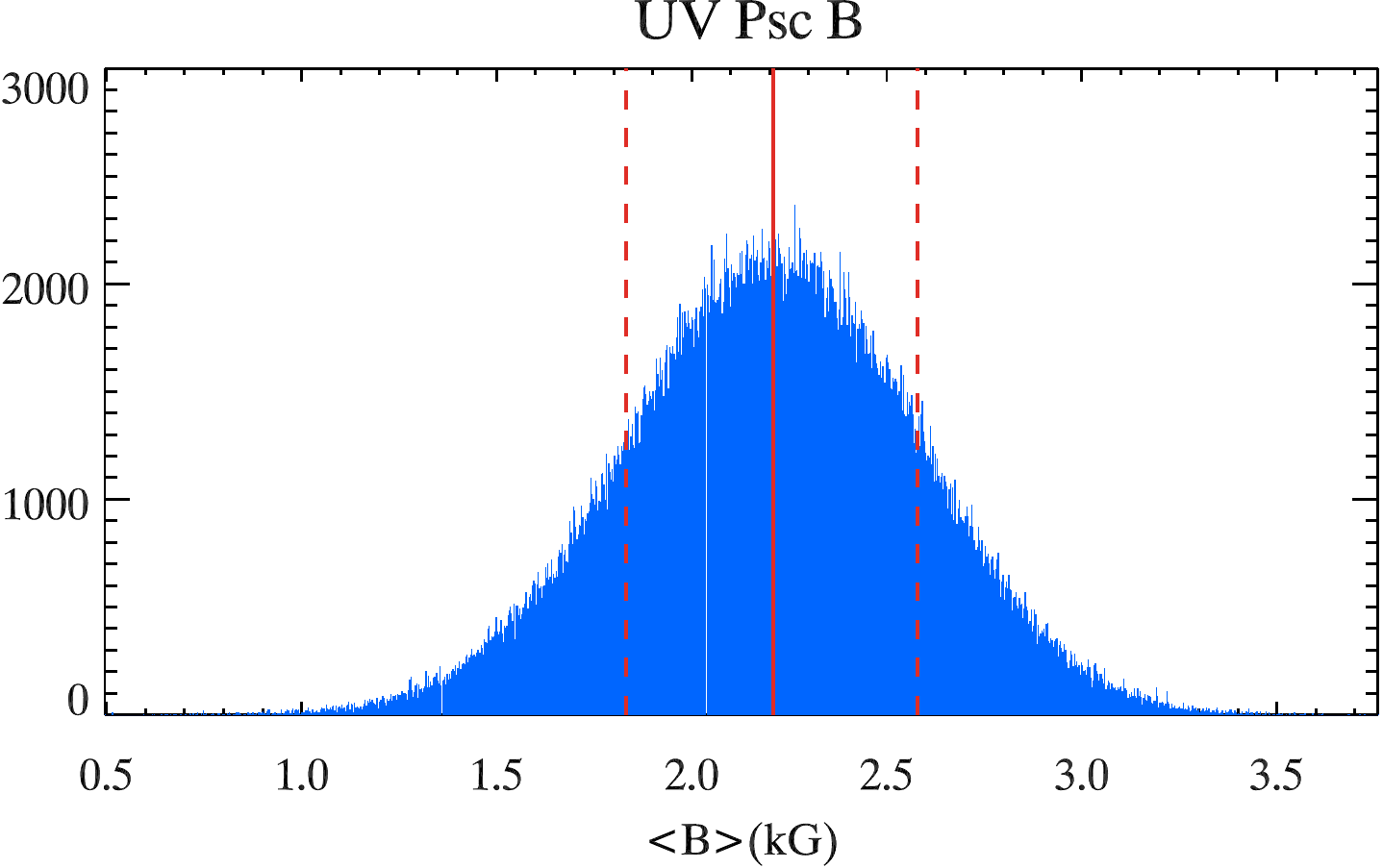}
  \caption{Result of the MCMC Zeeman intensification analysis for UV Psc B. The format of the plot is the same as in Fig.~\ref{fig:cornerA}, but the Fe abundance is replaced with the luminosity ratio.}
  \label{fig:cornerB}
\end{figure}

\begin{figure}
  \centering
  \includegraphics[width=\hsize]{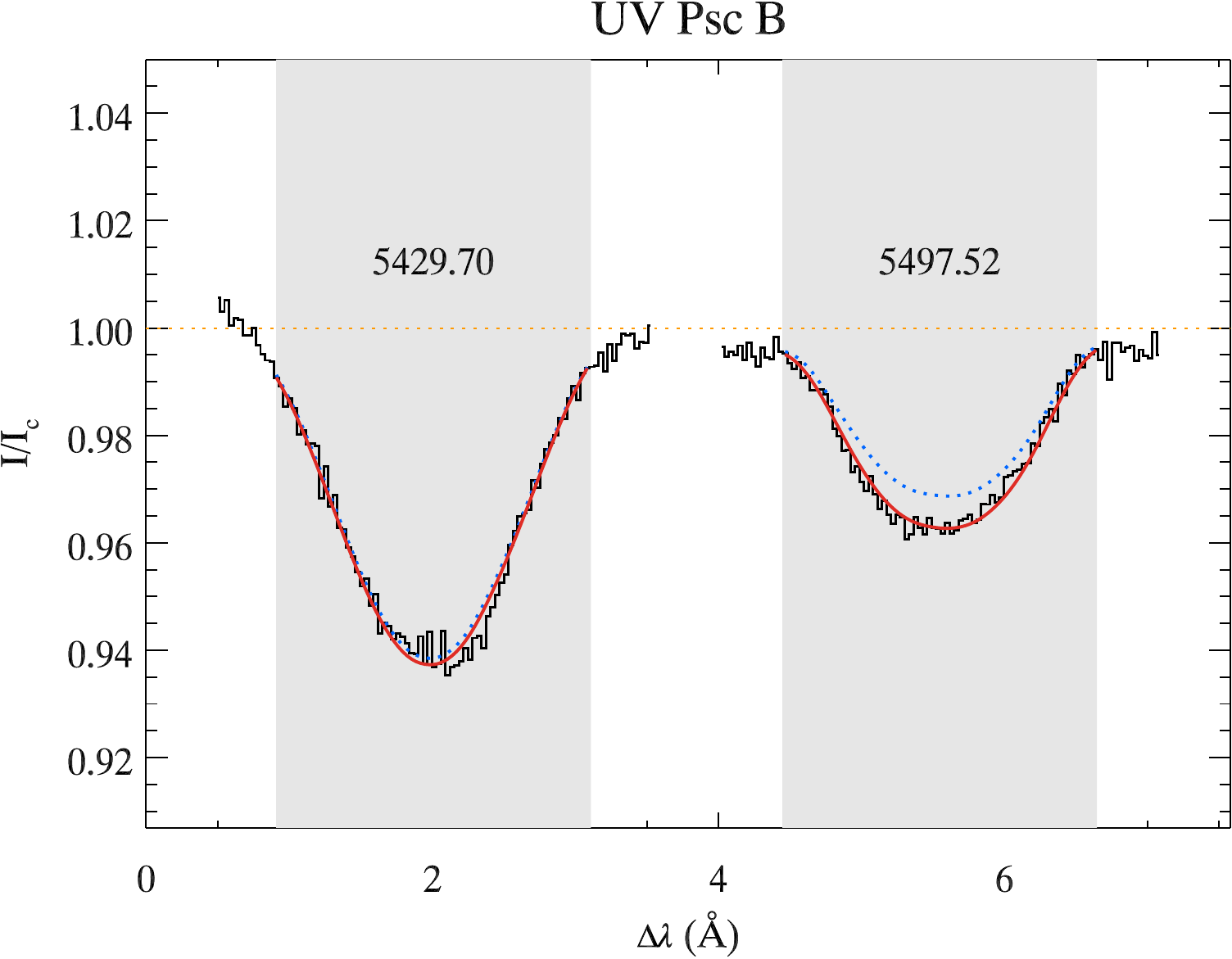}
  \caption{Same as Fig.~\ref{fig:line_fit}, but for UV Psc B.
}
  \label{fig:line_fitb}
\end{figure}

The results of the MCMC sampling for UV Psc B can be seen in Fig.~\ref{fig:cornerB} with the median parameter fit to observations shown in Fig.~\ref{fig:line_fitb}. Calculating the average magnetic field with the same procedure as was used for UV Psc A gives a magnetic field strength of $2.21\substack{+0.37\\-0.38}$~kG. 

These results show that Zeeman intensification can be utilised to measure the mean field strength even for rapidly rotating stars. While the high rotational velocity limits the line selection and causes a stronger degeneracy between contributions of individual \cla{field} components compared to stars rotating at a slower velocity, the method still provides a valuable insight into the average surface magnetic properties. \cla{While the average magnetic field strengths of the two components are similar, the distributions over field strengths are different. Figure \ref{fig:cornerA} indicates that UV Psc A is primarily covered by spots with strengths around 2\,kG. On the other hand, Fig. \ref{fig:cornerB} suggests that UV Psc B has predominantly 4~kG spots covering a smaller part of the surface. Although the uncertainties are large, these results could indicate a systematic difference in the field strength distribution for the small-scale fields on the two components.}

\cla{The stellar parameters of UV Psc are summarised in Table~\ref{tab:my_label}. The obtained values for $v_{\rm e}\sin{i}$ of the two components can be compared with the values calculated from radii and orbital period. The value for the primary is found to be between $65.1\pm1.4$ and $66.6\pm1.4$ km\,s$^{-1}$, depending on which equatorial radius is used (see Table~\ref{tab:rocheRadius}), compared to $71.0\pm0.4$ km\,s$^{-1}$ obtained from the MCMC inference. The difference is similar for the secondary (from $49.0\pm1.1$ to $49.9\pm1.1$ km\,s$^{-1}$ predicted, $51.7\pm1.2$ km\,s$^{-1}$ derived), although less significant given the larger error bars. This shows that the values obtained from the MCMC inference are larger than expected. Accounting for the Roche lobe geometry can mitigate but not fully account for this discrepancy. As mentioned earlier, this study does not account for other sources of non-magnetic broadening, which means that the $v_{\rm e}\sin{i}$ obtained in the MCMC inference is likely to be overestimated.}

\section{Summary and discussion}
\label{sec:summary}
\begin{table}
\cla{
    \caption{Stellar parameters for the two components of UV Psc}
    \label{tab:my_label}
    \begin{tabular}{lrr}
      \hline \hline
      Parameter & Primary & Secondary \\
      \hline
      Mass ($M_{\sun}$) & $1.0225 \pm 0.0058$  & $0.7741\pm0.0034$ \\
      Radius$^\mathrm{a}$ ($R_{\sun}$) & $1.110\pm0.023$ & $0.835\pm0.018$ \\
      $T_{\mathrm{eff}}$\,$^{\mathrm{b}}$ (K) & $5780\pm100$ & $4750\pm80$ \\
      $v_{\rm e}\sin{i}$ (km\,s$^{-1}$) & $71.01\substack{+0.37\\-0.36}$ & $51.66\substack{+1.25\\-1.20}$\\
      \hline
    \end{tabular}
    \tablefoot{(a) from \cite{torres:2010}, (b) from \cite{popper:1997}}
}
\end{table}
Our magnetic field analysis of UV Psc provided information on magnetic fields at different spatial scales in both components. By using the complementary techniques ZDI and Zeeman intensification, a more complete picture of magnetism in the system could be obtained. The results of our Zeeman intensification analysis indicated that this method can provide useful results for stars with significantly higher rotational velocities ($v_{\rm e}\sin i$ up to $\approx$\,70~km\,s$^{-1}$) compared to \cla{the} 30--40~km\,s$^{-1}$ limit achieved by previous studies \citep{kochukhov:2017,shulyak:2019}. This opens up the possibility to perform direct measurements of \cla{the} average magnetic field strengths on most cool active stars, including rapid rotators with periods below 1~d.

Comparing the average magnetic field strengths obtained from the two field diagnostic methods, the fractional field strength $\langle B_{V}\rangle/\langle B_{I}\rangle$ is $5.5$\% and $4.0$\% for UV Psc A and B respectively. In other words, the global field structures observed with the ZDI technique carries only about $0.3$\% and $0.1$\%, respectively of the magnetic energy that is contained in the small-scale fields detected with Zeeman intensification. This confirms that most magnetic energy is not visible to Stokes $V$ observations due to cancellation effects. The findings are also consistent with other sun-like stars studied by \cite{kochukhov:2020a}, showing similar fractions of magnetic field strengths recovered with ZDI. 

Other correlations between the properties of global magnetic fields and stellar parameters have been investigated. For example, \citet{vidotto:2014} obtained a relationship between the mean radial magnetic field strength inferred by ZDI studies and the rotation period of the star $\langle|B_{V}|\rangle\propto P_{\mathrm{rot}}^{-1.32\pm0.14}$. Both components of UV Psc agree well with this relationship. It is also of interest to see how the magnetic energy is distributed between the different types (poloidal and toroidal as well as axisymmetric and non-axisymmetric) of the harmonic field components, reported for UV Psc in Table \ref{tab:energyDist}. \citet{see:2015} investigated how these components relate to each other and stellar parameters. While large variations between different stars and epochs were found, certain parts of the parameter space appear to be avoided by real stars. One example is the relationship between the toroidal and axisymmetric ($m=0$) magnetic energies (fig.~6 in \citealt{see:2015}), with all stars with a substantial toroidal field contribution showing predominantly axisymmetric fields. UV Psc A is marginally consistent with this trend, falling on the border of the distribution formed by other ZDI results. This could also be due to the hemispheric degeneracy discussed in Sect. \ref{sec:degeneracy}. As we have established earlier, this degeneracy leads to underestimation of the strength of axisymmetric field components that are anti-symmetric with respect to the stellar equator, such as a dipole aligned with the rotational axis.

We found that the two observations done in early August 2016 could not be phased with the observations collected in September of the same year. This implies that the surface magnetic field structure has likely changed on the time scale of 1.5 months, resulting in different Stokes profiles observed at the same rotational phase. This highlights the fact that late-type active stars have dynamic and constantly changing surfaces. The surface field evolution can be caused by the differential rotation or emergency and decay of active regions -- with the data available for UV Psc we cannot distinguish between these two hypotheses. Similar to most current stellar magnetic field studies we relied on observations collected at only a single epoch. To gain a better understanding of the variations that occur on time scales of months and years a more systematic, multi-epoch magnetic monitoring \citep[e.g.][]{jeffers:2017,boro-saikia:2018} of selected systems is necessary.

Fundamental parameters of UV Psc were compared with stellar evolutionary models in several studies. There is a consensus that \cla{the} radii of both components are inflated compared to standard model predictions. The origin of this effect has been investigated by \citet{feiden:2013}. They included effects of magnetic suppression of convection in their stellar structure models in an attempt to reproduce the observed radii and arrive at a consistent age for the primary and secondary. Here we are able to test these predictions through direct field strength measurements. There is some tension between the field strengths obtained from observations and those adopted in theoretical models. The average field strength obtained from Zeeman intensification ($2.5\pm0.1$~kG) is slightly stronger for UV Psc A compared to the surface field intensity of 2.0\,kG required by the models. At the same time, our measured magnetic field strength for UV Psc B ($2.2\pm0.4$~kG) is weaker by a factor of about two compared to the 4.6\,kG field required by \citet{feiden:2013}. Possible mechanisms to reduce magnetic field strengths while still yielding correct stellar radii inflation were explored in the latter paper. Our results suggest the need for further evaluation of these processes. Such work would represent an important step in understanding the impact of magnetic fields on \cla{the} interior structure of active stars.

A common indirect method to obtain an estimate of magnetic field strength in a cool star is to relate it to the X-ray luminosity. This has been done for UV Psc by \cite{feiden:2013}, who used the ROSAT All-Sky Survey \citep{voges:1999} measurement of the system to obtain an estimate of the field and compare with the field strengths required by their models. Our results are in closer agreements with the indirect estimate from X-ray observations ($0.79^{+1.43}_{-0.51}$\,kG and $1.39^{+2.53}_{-0.90}$\,kG for UV Psc~A and B respectively) compared to the field strengths required by theoretical stellar interior models, with UV Psc B falling well within the uncertainties. UV Psc A does however have a field that is stronger than the observed X-ray luminosity would imply. This disagreement can be related to temporal variation in the magnetic activity of the star. Another possible source of this discrepancy could come from the way the magnetic fields were calculated from the X-ray luminosity by \citet{feiden:2013}. The X-ray flux is an estimate of the flux per star, calculated by dividing the total observed flux of the system by the number of stars contributing to the flux. While there are no other sources contaminating the X-ray observations of UV Psc, it means that UV Psc A and B were assumed to have the same X-ray luminosity and hence an equal magnetic flux, \cla{an} assumption that may be inaccurate. Our finding of \cla{a stronger magnetic field on the} primary component compared to the secondary\cla{, combined with the fact that the primary has a larger surface area, indicates that magnetic fluxes are not equal.}

A general trend within this work has been that the errors and uncertainties are greatly amplified for UV Psc B. This is to be expected since it is a smaller and fainter component in the system. But it still highlights a general issue with studying binary systems where one star is significantly brighter compared to the other one. 

In this study we have systematically assessed, for the first time, the impact of hemisphere degeneracy on the ZDI maps of stars observed close to the equator. When studying synthetic field structures applied to the surface\cla{s} of \cla{the} UV Psc components it was found that this system suffers from a large-scale polarisation signal cancellation. This is in addition to the small-scale signal cancellation present in all ZDI studies. Investigation of \cla{the} polarisation signatures of individual spherical harmonic modes showed that those modes that were antisymmetric with respect to the equator were cancelled out by this effect. This indicates that magnetic field studies only relying on circular polarisation will inevitably underestimate the overall strength of the global field structures on stars with an inclination close to $90^{\circ}$.

In the near future, spectropolarimetric observations using high-resolution near-infrared spectrometers SPIRou \citep{donati:2020} and CRIRES$^+$ \citep{dorn:2016} could further improve the opportunities for magnetic field studies based on complementary ZDI and Zeeman broadening or intensification approaches. Furthermore, obtaining linear polarisation observations for cool stars similar to UV Psc could help \cla{to solve} many of the issues encountered when applying the ZDI method to cool stars in general and eclipsing binaries in particular. This includes the ability to more accurately recover high-order components of the global magnetic field structure as well as reducing the systematic underestimation of the meridional field component \citep{rosen:2012,rosen:2015}. Furthermore, our study shows that linear polarisation observations can be used to detect antisymmetric magnetic field structures, thus lifting the hemisphere degeneracy for perpendicular rotators. Even individual detections of the Zeeman-induced linear polarisation in spectral lines can be helpful in ascertaining the presence of simple axisymmetric fields in eclipsing binaries observed equator-on. These considerations also apply to the majority of transiting exoplanet host stars, which tend to have aligned rotational and orbital axes \citep{campante:2016,winn:2017,munoz:2018}. For example, \citet{kochukhov:2020b} inferred the presence of a strong axisymmetric dipolar field from a few serendipitous linear polarisation observations of the young, very active M-dwarf exoplanet host AU Mic. In contrast, this field component is missing from the magnetic maps recovered with conventional ZDI using Stokes $V$ data \citep{kochukhov:2020b,klein:2021}. The same bias is likely to be present in published ZDI analyses of other magnetically active transiting exoplanet hosts \citep[e.g.][]{fares:2010}, whose global magnetic fields may thus be significantly underestimated.

\begin{acknowledgements}
We thank Frida Torr{\aa}ng for her contribution to the initial analysis of UV Psc data, the CFHT staff for their assistance throughout the observing runs \cla{and the referee for their careful reading and insightful comments}. O.K. acknowledges support by the Swedish Research Council, the Swedish National Space Agency, and the Royal Swedish Academy of Sciences. Based on observations obtained at the CFHT which is operated by the National Research Council of Canada, the Institut National des Sciences de l'Univers (INSU) of the Centre National de la Recherche Scientifique (CNRS) of France, and the University of Hawaii. \cla{This work has made use of the VALD database, operated at Uppsala University, the Institute of Astronomy RAS in Moscow, and the University of Vienna.}
\end{acknowledgements}

%

%

\begin{appendix}

\section{Radial velocity measurements}
\begin{table}[H]
    \centering
    \caption{Radial velocities of UV Psc components.}
    \label{tab:RV}
    \begin{tabular}{cccc|cccc}
\hline
\hline
Reduced HJD & S/N$_{I}$ & $V_{A}$ & $V_{B}$ & Reduced HJD & S/N$_{I}$ & $V_{A}$ & $V_{B}$\\
&  & (km\,s$^{-1}$) & (km\,s$^{-1}$) & & & (km\,s$^{-1}$) & (km\,s$^{-1}$) \\
\hline
57604.1047 & 160 & 122.36 & -154.97 & 57651.8413 & 156  & -102.84 &  152.11 \\
57604.1104 & 162 & 123.70 & -148.03 & 57651.8470 & 158 & -104.77 &  154.58 \\
57604.1161 & 162 & 123.59 & -147.86 & 57651.8527 & 156 & -106.60 &  156.83 \\
57604.1217 & 161 & 123.31 & -147.53 & 57651.8584 & 162 & -108.14 &  158.48 \\
57607.1256 & 128 & -113.69 &  165.15 & 57651.9693 & 155 & -98.08 &  143.62 \\
57607.1313 & 122 & -113.50 &  164.89 & 57651.9749 & 149 & -95.57 &  140.41 \\
57607.1370 & 124 & -113.07 &  164.38 & 57651.9806 & 149 & -92.66 &  136.71 \\
57607.1427 & 131 & -112.48 &  163.72 & 57651.9863 & 149 & -89.71 &  132.89 \\
57649.8811 & 104 & 68.79 &  -79.72 & 57652.0830 & 143 & -25.59 &   46.29 \\
57649.8868 & 114 & 65.55 &  -75.00 & 57652.0887 & 141 & -21.45 &   38.51 \\
57649.8925 & 119 & 61.59 &  -70.00 & 57652.0944 & 140 & -17.05 &   32.19 \\
57649.8982 & 120 & 57.42 &  -64.53 & 57652.1001 & 135 & -11.96 &   26.84 \\
57649.9045 & 130 & 52.71 &  -58.04 & 57652.8214 & 156 & -101.74 &  148.88 \\
57649.9102 & 123 & 49.93 &  -57.43 & 57652.8271 & 154 & -99.44 &  145.73 \\
57649.9159 & 118 & 48.09 &  -57.45 & 57652.8328 & 153 & -96.93 &  142.25 \\
57649.9216 & 121 &  46.60 &  -55.97 & 57652.8385 & 155 &  -94.29 &  138.79 \\
57650.0112 & 155 & -38.14 &   74.15 & 57653.0047 & 141 & 36.63 &  -36.21 \\
57650.0169 & 153 & -40.31 &   76.42 & 57653.0104 & 144 & 40.59 &  -38.38 \\
57650.0226 & 156 & -44.29 &   80.45 & 57653.0161 & 148 & 43.71 &  -40.71 \\
57650.0283 & 157 & -48.38 &   85.54 & 57653.0217 & 151 & 46.97 &  -43.06 \\
57650.1215 & 161 & -104.26 &  152.02 & 57653.0754 & 154 & 84.06 &  -93.48 \\
57650.1272 & 161 & -106.04 &  154.59 & 57653.0811 & 156 & 87.64 &  -97.75 \\
57650.1329 & 160 & -107.73 &  156.82 & 57653.0868 & 146 & 91.11 & -102.24 \\
57650.1386 & 159 & -109.19 &  158.64 & 57653.0925 & 149 & 94.32 & -106.78 \\
57650.8752 & 156 & -40.30 &   71.70 & 57653.9370 & 160 & 84.31 &  -93.80 \\
57650.8809 & 161 & -43.57 &   75.15 & 57653.9427 & 163 & 87.80 &  -98.41 \\
57650.8866 & 161 & -48.01 &   80.57 & 57653.9483 & 162 & 91.26 & -102.59 \\
57650.8923 & 162 & -51.97 &   86.37 & 57653.9540 & 160 & 94.48 & -107.06 \\
57650.9637 & 165 & -95.91 &  143.28 & 57654.0107 & 161 & 118.05 & -140.72 \\
57650.9694 & 162 &  -98.51 &  146.39 & 57654.0164 & 161 & 119.31 & -142.68 \\
57650.9751 & 161 & -100.82 &  149.33 & 57654.0221 & 160 & 120.43 & -144.11 \\
57650.9808 & 160 & -103.00 &  152.32 & 57654.0278 & 164 & 121.38 & -145.32 \\
57651.0604 & 158 & -111.88 &  162.24 & 57654.1024 & 154 & 115.25 & -137.37 \\
57651.0661 & 163 & -111.07 &  161.19 & 57654.1081 & 159 & 113.33 & -134.65 \\
57651.0718 & 159 & -110.01 &  159.60 & 57654.1138 & 163 & 111.32 & -131.58 \\
57651.0775 & 162 & -108.65 &  157.81 & 57654.1195 & 162 & 109.04 & -128.49 \\
\hline
    \end{tabular}
\end{table}

\section{Magnetic reconstruction at lower inclinations}
\begin{figure*}
    \centering
    \includegraphics[width=0.5\textwidth]{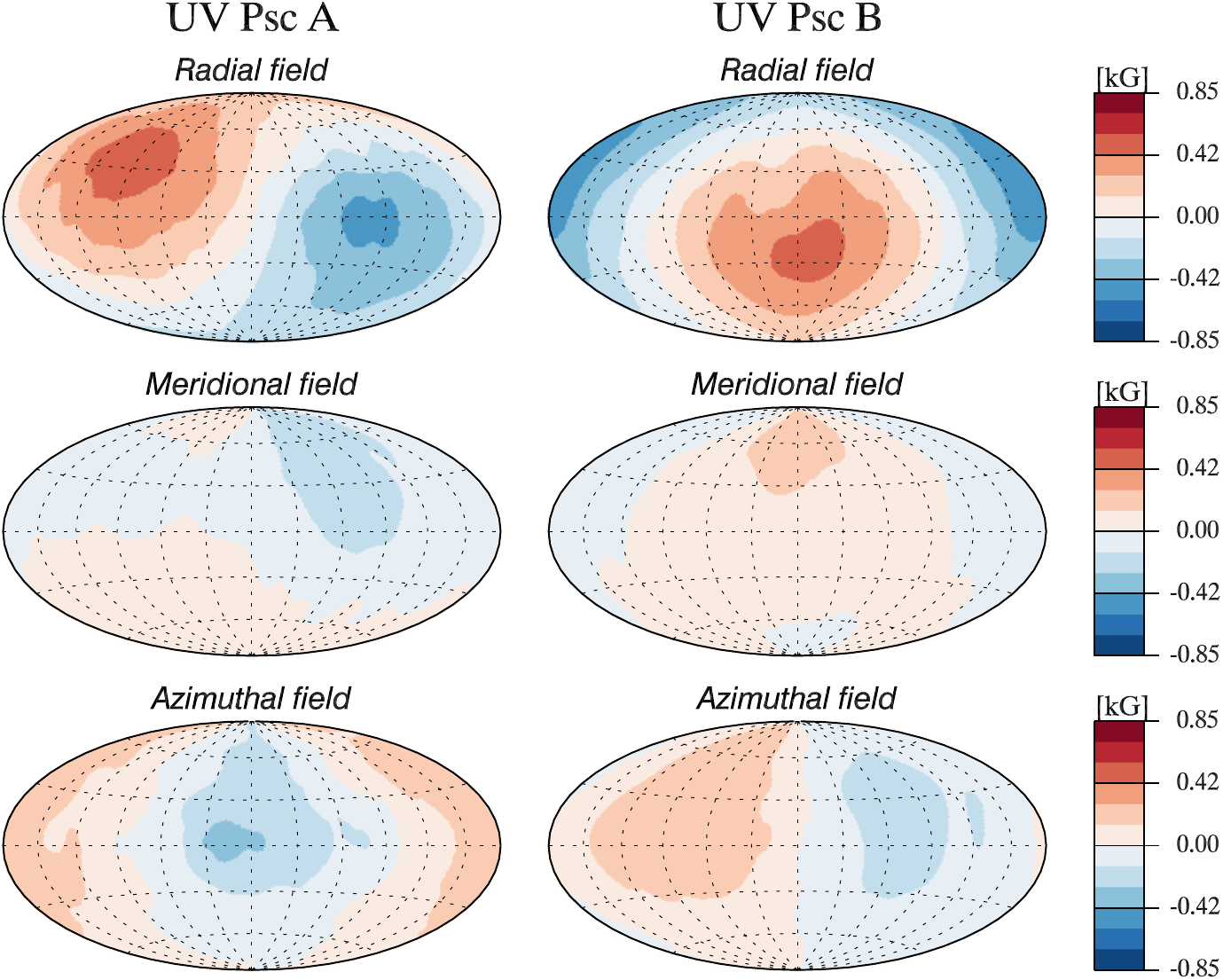}\includegraphics[width=0.5\textwidth]{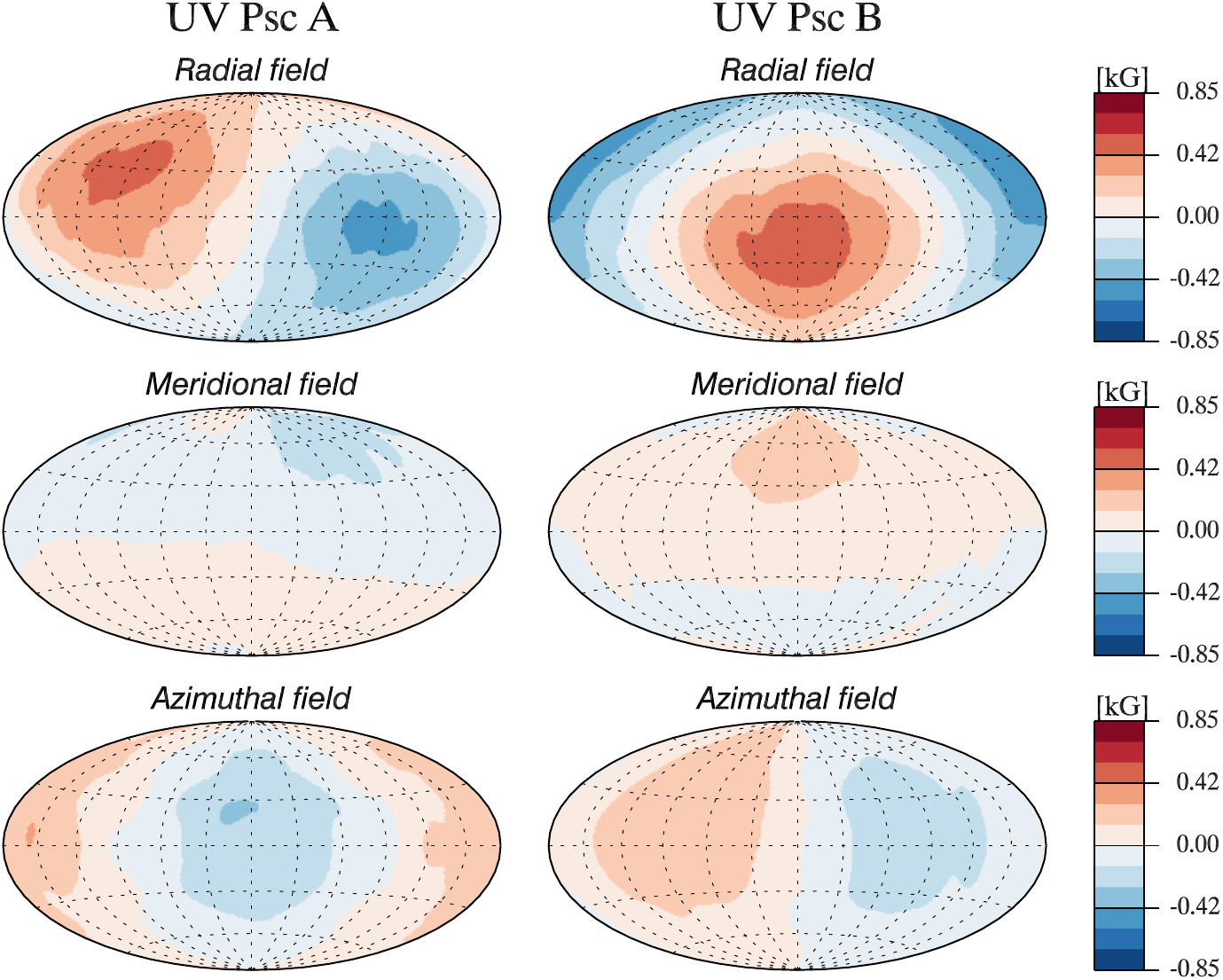}
    \caption{\cla{Magnetic field reconstruction performed on the same magentic geometry as in Fig. \ref{fig:synB} but with different system inclination. \textit{Left.} Inclination of 80$^\circ$. \textit{Rigth.} Inclination of 85$^\circ$}}
    \label{fig:SynBInclination}
\end{figure*}
\end{appendix}
\end{document}